\pdfoutput=1

\documentclass[11pt,twoside,a4paper,cmspaper,final,collab]{cms-tdr}
\def\svnVersion{364521}\def\cmsCernNoTag{CERN-EP/2016-104}\def\cmsCernDate{\today}\def\cmsMessage{Published in the European Physical Journal C as \href{http://dx.doi.org/10.1140/epjc/s10052-016-4286-3}{\doi{10.1140/epjc/s10052-016-4286-3.}}}

\begin{document}\cmsNoteHeader{SMP-15-007}

\hyphenation{had-ron-i-za-tion}
\hyphenation{cal-or-i-me-ter}
\hyphenation{de-vices}
\RCS$Revision: 354371 $
\RCS$HeadURL: svn+ssh://svn.cern.ch/reps/tdr2/papers/SMP-15-007/trunk/SMP-15-007.tex $
\RCS$Id: SMP-15-007.tex 354371 2016-07-10 17:39:43Z alverson $
\newlength\cmsFigWidth
\ifthenelse{\boolean{cms@external}}{\setlength\cmsFigWidth{0.85\columnwidth}}{\setlength\cmsFigWidth{0.4\textwidth}}
\ifthenelse{\boolean{cms@external}}{\providecommand{\cmsLeft}{top\xspace}}{\providecommand{\cmsLeft}{left\xspace}}
\ifthenelse{\boolean{cms@external}}{\providecommand{\cmsRight}{bottom\xspace}}{\providecommand{\cmsRight}{right\xspace}}
\providecommand{\alps}{\ensuremath{\alpha_S}\xspace}
\providecommand{\alpsmz}{\ensuremath{\alpha_S(M_{\PZ})}\xspace}
\providecommand{\NLOJETPP} {{\textsc{NLOJet++}}\xspace}
\providecommand{\fastNLO} {{\textsc{fastNLO}}\xspace}
\providecommand{\fastjet} {{\textsc{FastJet}}\xspace}
\providecommand{\PYTHIAE} {{\textsc{pythia8}}\xspace}
\providecommand{\RooUnfold} {{\textsc{RooUnfold}}\xspace}
\providecommand{\kts}{\ensuremath{k_{\mathrm{t}}}\xspace}
\providecommand{\mur}{\ensuremath{\mu_r}\xspace}
\providecommand{\muf}{\ensuremath{\mu_f}\xspace}

\cmsNoteHeader{SMP-15-007}
\title{Measurement of the double-differential inclusive jet cross section in
 proton-proton collisions at $\sqrt{s} = 13\TeV$}

\date{\today}

\abstract{A measurement of the double-differential inclusive jet cross
  section as a function of jet transverse momentum $\pt$ and
  absolute jet rapidity $\abs{y}$ is presented. The
  analysis is based on proton-proton collisions collected by the CMS
  experiment at the LHC at a centre-of-mass energy of 13\TeV. The data samples
  correspond to integrated luminosities of~71
  and~44\pbinv for $\abs{y}<3$ and $3.2<\abs{y}<4.7$, respectively.
  Jets are reconstructed with the anti-$\kts$ clustering algorithm for two jet
  sizes, $R$, of 0.7 and 0.4, in a phase space region covering jet
  $\pt$ up to 2\TeV and jet rapidity up to $\abs{y}$ = 4.7.
  Predictions  of perturbative quantum chromodynamics at next-to-leading order precision,
  complemented with electroweak and nonperturbative corrections,
  are used to compute the absolute scale and the shape
  of the inclusive jet cross section. The cross
  section difference in $R$, when going to a smaller jet size of~0.4,
  is best described by Monte Carlo event generators with next-to-leading order
  predictions matched to parton showering, hadronisation, and
  multiparton interactions. In the phase space accessible with the new data,
  this measurement provides a first indication that jet physics is as well understood at
  $\sqrt{s}=13\TeV$ as at smaller centre-of-mass energies.}

\hypersetup{%
pdfauthor={CMS Collaboration},%
pdftitle={Measurement of the double-differential inclusive jet cross section in
 proton-proton collisions at sqrt(s)=13 TeV},%
pdfsubject={CMS},%
pdfkeywords={CMS, physics, QCD}}

\maketitle

\section{Introduction}
\label{sec:intro}

Quantum chromodynamics (QCD) is the fundamental theory describing
strong interactions among partons, \ie quarks and gluons. Inclusive jet production
($\Pp+\Pp\to \text{jet}+ \mathrm{X}$) is a key process
to test predictions of perturbative QCD (pQCD) over a wide region in
phase space. To compare with measurements, the parton-level calculations must
be complemented with corrections for nonperturbative (NP) effects that
involve the modeling of hadronisation (HAD) and multiparton
interactions (MPI). Previous measurements at the CERN LHC have been
carried out by the ATLAS and CMS Collaborations at centre-of-mass energies $\sqrt{s} = 2.76\TeV$~\cite{Aad:276TeV, Khachatryan:2015luy},
7\TeV~\cite{Aad:2010wv, CMSInclusive, Aad:2011fc, Chatrchyan:2012bja, Aad:2014vwa}, and at lower $\sqrt{s}$ by experiments at other hadron
colliders~\cite{Banner:1982kt, Arnison:1983gw, Abulencia:2007ez,
  Abazov:2008hu, Aaltonen:2008eq}.
The measurements at 2.76 and 7\TeV centre-of-mass energies were found to
be in agreement with calculations at next-to-leading order (NLO) in
the strong coupling constant $\alpha_S$ over a wide range of jet
transverse momentum \pt and rapidity $y$.
With the latest data from the LHC Run 2, these tests of pQCD are
extended to cover the new energy regime of $\sqrt{s}=13\TeV$.

In this paper, a measurement of the double-differential inclusive jet
cross section is presented as a function of the jet \pt and
absolute jet rapidity $\abs{y}$. The jets are clustered with the anti-\kts
jet algorithm~\cite{Cacciari:2008gp} as implemented in the \fastjet
library~\cite{Cacciari:2011ma}. Two jet sizes $R$ are used: the larger
value $R=0.7$ corresponds to the standard jet size chosen in most QCD
jet analyses made by the CMS Collaboration because it favourably
compares to fixed-order predictions~\cite{Chatrchyan:2014gia}.
A second, smaller value of $R$ emphasizes different aspects of
perturbative and nonperturbative QCD and permits complementary tests
to be performed~\cite{Dasgupta:2007wa, Dasgupta:2014yra,
  Dasgupta:2016bnd}.  Moreover, the choice of $R=0.4$ as a new CMS
default jet size that replaces the previous one of~0.5 in LHC Run~1
analyses will allow direct comparisons between jet measurements made
by ATLAS and CMS\@.

The proton-proton collision data were recorded by the CMS experiment
at a centre-of-mass energy of 13\TeV in 2015. The data samples correspond to
integrated luminosities of~71 and 44\pbinv for ranges in rapidity of
$\abs{y} < 3$ and $3.2 < \abs{y} < 4.7$, respectively. The smaller amount of
data for the forward rapidity range is explained by more difficult
operating conditions at the very start of data taking, which reduced the event sample certified for physics analyses.
The results are compared to fixed-order
predictions at NLO precision, complemented with electroweak and
nonperturbative corrections, and to predictions of various Monte Carlo
(MC) event generators that combine leading-order (LO) or NLO pQCD with
the modeling of parton showers (PS), HAD, and MPI.

\section{The CMS detector}
\label{sec:detector}

The central feature of the CMS apparatus is a superconducting solenoid
of 6\unit{m} internal diameter, providing a magnetic field of
3.8\unit{T}. Within the solenoid volume are a silicon pixel and strip
tracker, a lead tungstate crystal electromagnetic calorimeter (ECAL),
and a brass and scintillator hadron calorimeter (HCAL), each composed
of a barrel and two endcap sections. Forward calorimeters extend the
pseudorapidity ($\eta$) coverage provided by
the barrel and endcap detectors to the region $3.0 < \abs{y} < 5.2$. Muons are measured in gas-ionisation
detectors embedded in the steel flux-return yoke outside the solenoid.
In the region $\abs{\eta} < 1.74$, the HCAL cells have widths of
0.087 in $\eta$ and 0.087 radians in azimuth ($\phi$). In the
$\eta$-$\phi$ plane, and for $\abs{\eta} < 1.48$, the HCAL cells map
onto $5 \times 5$ ECAL crystals arrays to form calorimeter towers
projecting radially outwards from close to the nominal interaction
point. At larger values of $\abs{\eta}$, the size in rapidity of the towers
increases and the matching ECAL arrays contain fewer crystals. Within
each tower, the energy deposits in ECAL and HCAL cells are summed to
define the calorimeter tower energies, subsequently used to provide
the energies and directions of hadronic jets.
The particle-flow (PF) event algorithm~\cite{CMS:2009nxa,CMS:2010byl}
reconstructs and identifies each individual particle with an optimised
combination of information from the various elements of the CMS
detector. The energy of photons is directly obtained from the ECAL
measurement. The energy of electrons is determined from a combination
of the electron momentum at the primary interaction vertex as
determined by the tracker, the energy of the corresponding ECAL
cluster, and the energy sum of all bremsstrahlung photons spatially
compatible with originating from the electron track. The momentum of
muons is obtained from the curvature of the corresponding track. The
energy of charged hadrons is determined from a combination of their
momenta measured in the tracker and the matching ECAL and HCAL energy
deposits, corrected for zero-suppression effects and for the response
function of the calorimeters to hadronic showers. Finally, the energy
of neutral hadrons is obtained from the corresponding ECAL
and HCAL energy. When combining information from the entire detector,
the jet energy resolution typically amounts to 15\% at 10\GeV, 8\% at
100\GeV, and 4\% at 1\TeV, to be compared to about 40\%, 12\%, and 5\%
obtained when the ECAL and HCAL alone are used.  A more detailed
description of the CMS detector, together with a definition of the
coordinate system used and the relevant kinematic variables, can be
found in Ref.~\cite{Chatrchyan:2008zzk}.

\section{Event selection and jet reconstruction}
\label{sec:selectionReconstruction}

The measurement is based on data samples collected with single-jet
high-level triggers (HLT)~\cite{CMS_HLT}.  Eight single-jet HLT paths are
considered, seeded by Level~1 triggers based on calorimetric
information.  They require, in the full rapidity coverage of the CMS
detector, at least one jet in each event with $\pt > 60$, 80,
140, 200, 260, 300, 400, or 450\GeV. All triggers, except the
one with the highest threshold, are prescaled.
The relative efficiency of each trigger is estimated using
lower-\pt-threshold triggers, and found to exceed 99\% in the
\pt regions shown in Table~\ref{TriggerRegions}. The absolute trigger
efficiency is measured using a tag and probe
method~\cite{Chatrchyan:2011ds} based on events selected with a single-jet
trigger threshold of 40\GeV, a back-to-back dijet
system, and a probe jet matched to a HLT trigger object. This trigger has
an efficiency greater than 99\% for selecting an event with a jet
of $\pt > 80\GeV$.

\begin{table}[htbp]
  \begin{center}
    \begin{tabular}{c c}
      \hline
      HLT path &  \pt range (\GeVns)   \\
      \hline
      PFJet$\_$60  & 114--133      \\
      PFJet$\_$80  & 133--220      \\
      PFJet$\_$140 & 220--300      \\
      PFJet$\_$200 & 300--430      \\
      PFJet$\_$260 & 430--507      \\
      PFJet$\_$300 & 507--638      \\
      PFJet$\_$400 & 638--737      \\
      PFJet$\_$450 & ${>}737$      \\
      \hline
    \end{tabular}
    \caption{Trigger regions defined as ranges of the leading jet
      \pt in each event for all single-jet triggers used in
      the inclusive jet cross section measurement.}

    \label{TriggerRegions}
  \end{center}
\end{table}

The main physics objects in this analysis are PF jets, reconstructed by clustering the Lorentz vectors of the PF candidates
with the anti-\kts (AK) clustering algorithm for the two jet sizes $R
= 0.7$ and~0.4 that will be referred to as AK7 and AK4,
respectively.
 In order to reduce the contribution to the reconstructed
jets from additional proton-proton interactions within the same or
neighbouring bunch crossings (pileup), the technique of charged hadron
subtraction~\cite{CMS:2012woa} is used. Pileup produces
unwanted calorimetric energy depositions and additional tracks. The charged hadron subtraction
reduces these effects by removing char\-ged particles that originate from pileup
vertices.
The average number of pileup interactions observed in these
data is ${\approx}19$. During data collection the LHC operated with a $50\unit{ns}$ bunch spacing.

Reconstructed jets require small energy corrections to account for
residual nonuniformities and nonlinearities in the detector response.
Jet energy scale (JES)~\cite{Chatrchyan:2011ds} corrections are obtained using
simulated events, generated with \PYTHIAE.204~\cite{Sjostrand:2014zea}
with tune CUETM1~\cite{Khachatryan:2015pea} and processed through the
CMS detector simulation, and in situ measurements with dijet,
photon+jet, and $\PZ$+jet events.  An offset correction is applied to
account for the extra energy clustered into jets due to the contribution of neutral particles produced by additional
pileup interactions within the same or neighbouring bunch
crossings.

The JES correction, applied as a multiplicative factor to
the jet four-momentum vector, depends on the jet $\eta$ and \pt
values. The typical correction is about 10\% for a central jet with a
\pt of 100\GeV, and decreases with increasing \pt.

Events are required to have at least one primary vertex (PV). If more than one primary vertex is present, the vertex
with the highest sum of the squared \pt of the associated tracks is selected. This selected vertex is required to be reconstructed
from at least five charged-particle tracks and must satisfy a set of
quality requirements, including $\abs{z_\mathrm{PV}} < 24\unit{cm}$ and
$\rho_{\mathrm{PV}} < 2\unit{cm}$, where $z_{\mathrm{PV}}$ and
$\rho_{\mathrm{PV}}$ are the longitudinal and transverse distances of
the primary vertex from the nominal interaction point in
the CMS detector.
Jets with $\pt > 114\GeV$ are grouped in
seven different $\abs{y}$ bins.
Additional selection criteria are applied to each event to remove
spurious jet-like signatures originating from isolated noise patterns
in certain HCAL regions. To suppress noise patterns, tight
identification criteria are applied~\cite{CMS:JME-10}: each jet should contain at least
two particles, one of which is a charged hadron, and the jet energy
fraction carried by neutral hadrons and photons should be less than
90\%. These criteria have an efficiency greater than 99\% for genuine
jets.

\section{Measurement of the double-differential inclusive jet cross
  section}
\label{sec:measurement}

The double-differential inclusive jet cross section is defined as
\begin{equation}
  \label{master_inclusive}
  \frac{\rd ^2\sigma}{\rd \pt\rd y} =
  \frac{1}{\epsilon\lumi}\,\frac{N_\mathrm{j}}{\Delta\pt \Delta y},
\end{equation}
where \lumi is the integrated luminosity, $N_\mathrm{j}$ is the
number of jets in a bin of a width $\Delta \pt$ in transverse
momentum and $\Delta y$ in rapidity, and $\epsilon$ is the product of
the trigger and jet selection efficiencies, which is greater than
$99\%$. The phase space in rapidity is subdivided into six bins from
$y=0$ to $\abs{y}=3$ with $\abs{\Delta y} = 0.5$, and one bin from $\abs{y}=3.2$
to $4.7$, the forward rapidity region. The bin width in \pt
is chosen in such a way that bin-to-bin migrations due to detector resolution are less than 50\%. In each bin, the statistical uncertainty is derived through the formula $\sqrt{\smash[b]{(4-3f)/(2-f)}}\sqrt{\smash[b]{N_{\text{jets}}}}$, where $f$ corresponds to the fraction of events which contribute with exactly one jet in the bin \cite{Chatrchyan:2012bja}. This procedure corrects for
possible multiple entries per event. The fraction $f$ is typically larger than 95\% in the entire phase-space considered, thus the correction is small.

The double-differential inclusive jet cross section is corrected for
the detector resolution and unfolded to the stable particle level~\cite{Buttar:2008jx}. In
this way, a direct comparison of this measurement to results from other
experiments and to QCD predictions is possible. Particles are
considered stable if their mean path length $c\tau$ is greater than 10\unit{mm}.

The unfolding procedure is based on the iterative d'A\-gos\-ti\-ni
method~\cite{DAgostini:1994zf}, as implemented in the \RooUnfold
software package~\cite{bib:RooUnfold}, using a response matrix that
maps the predicted distribution onto the measured one. The response
matrix is derived from a simulation, that uses the theoretically
predicted spectrum as input and introduces smearing effects by taking
into account the jet \pt resolution.  The predicted spectrum is
evaluated from fixed-order calculations based on the \NLOJETPP v4.1.13
program~\cite{Nagy:2001fj,Nagy:2003tz} within the framework of the
\fastNLO v2.3.1 package~\cite{Britzger:2012bs}, using the
CT14~\cite{Dulat:2015mca} parton distribution functions (PDF). More
details are presented in Section~\ref{sec:fixedorder}. The jet
\pt resolution is evaluated with the CMS detector simulation
based on \GEANTfour~\cite{Agostinelli:2002hh} using a QCD simulation
from \PYTHIAE with tune CUETM1, after correcting for the residual
differences between data and simulation~\cite{Chatrchyan:2011ds}. The unfolded distributions differ
from the distributions at detector level by 5\% to $20\%$. The
unfolding procedure can turn statistical fluctuations of the measured
spectra into correlated patterns among the neighbouring bins.
It has been verified that such effects are always within the statistical
uncertainties of the unfolded distributions, which are larger
than those of the detector-level distributions. The iterative unfolding procedure
is regularized by limiting the number of iterations to four in each rapidity bin.

The main systematic uncertainties for the jet cross section
measurements arise from the JES calibration and from the uncertainty
in the integrated luminosity. The JES uncertainty, evaluated
separately for AK7 and AK4 jets, is 1--3\% in the central region
($\abs{y} < 2$) and increases to 7--8\% in the forward rapidity region
($3.2 < \abs{y} < 4.7$)~\cite{Chatrchyan:2011ds}. The JES uncertainty also includes the uncertainty
carried by the charged hadron subtraction.
The resulting uncertainties in the double-differential inclusive jet cross section
range between 8\% at central rapidities and low \pt to 65\% at
forward rapidities and the highest \pt. The uncertainty in the
integrated luminosity (2.7\% \cite{CMS:LUM-15-001}) propagates directly
to the cross section.

The unfolding procedure is affected by uncertainties in the jet energy
resolution (JER) parame-\\trisation. Alternative response matrices are
used to unfold the measured spectra. They are built by varying the JER
parameters within their uncertainties~\cite{Chatrchyan:2011ds}. The JER uncertainty
introduces a 1--2\% uncertainty in the measured cross section. The
model dependence of the theoretical \pt spectrum also affects
the response matrix and thus the unfolding, but this uncertainty has
negligible effects on the cross section measurement. The model
dependence is checked using various PDF sets to calculate the theoretical \pt
spectrum.

Finally, an uncertainty of 1\% is assigned to the cross section to
account for residual effects of small inefficiencies from jet
identification~\cite{Chatrchyan:2014gia}. The total experimental
systematic uncertainty of the measured cross section is obtained by
summing in quadrature the individual contributions from JES,
luminosity, JER, and jet identification uncertainties.

\section{Theoretical predictions}
\label{sec:theory}
\subsection{Predictions from fixed-order calculations in pQCD}
\label{sec:fixedorder}

\begin{figure*}[t]
  \centering
  \includegraphics[width=0.45\textwidth]{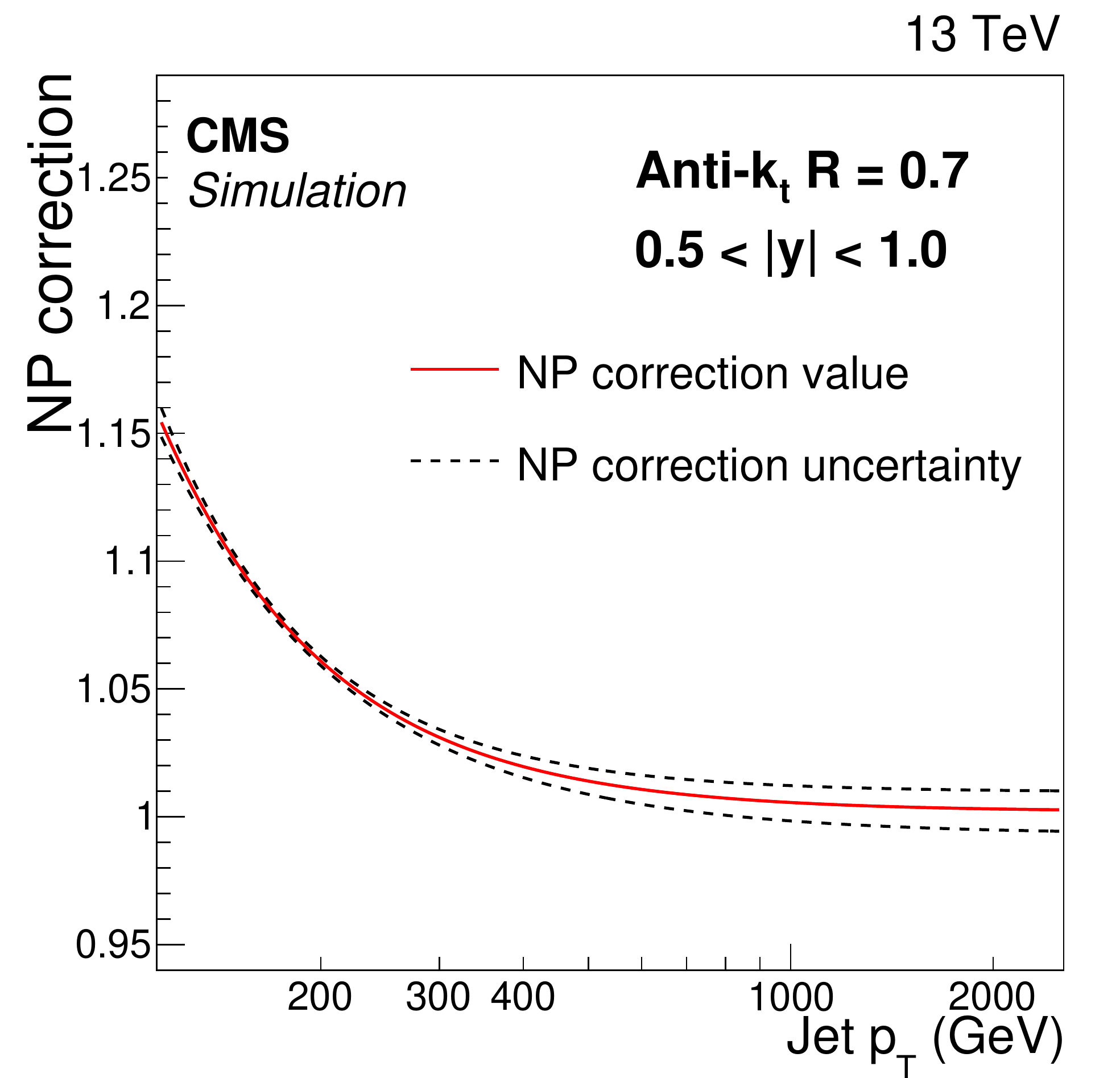}
  \includegraphics[width=0.45\textwidth]{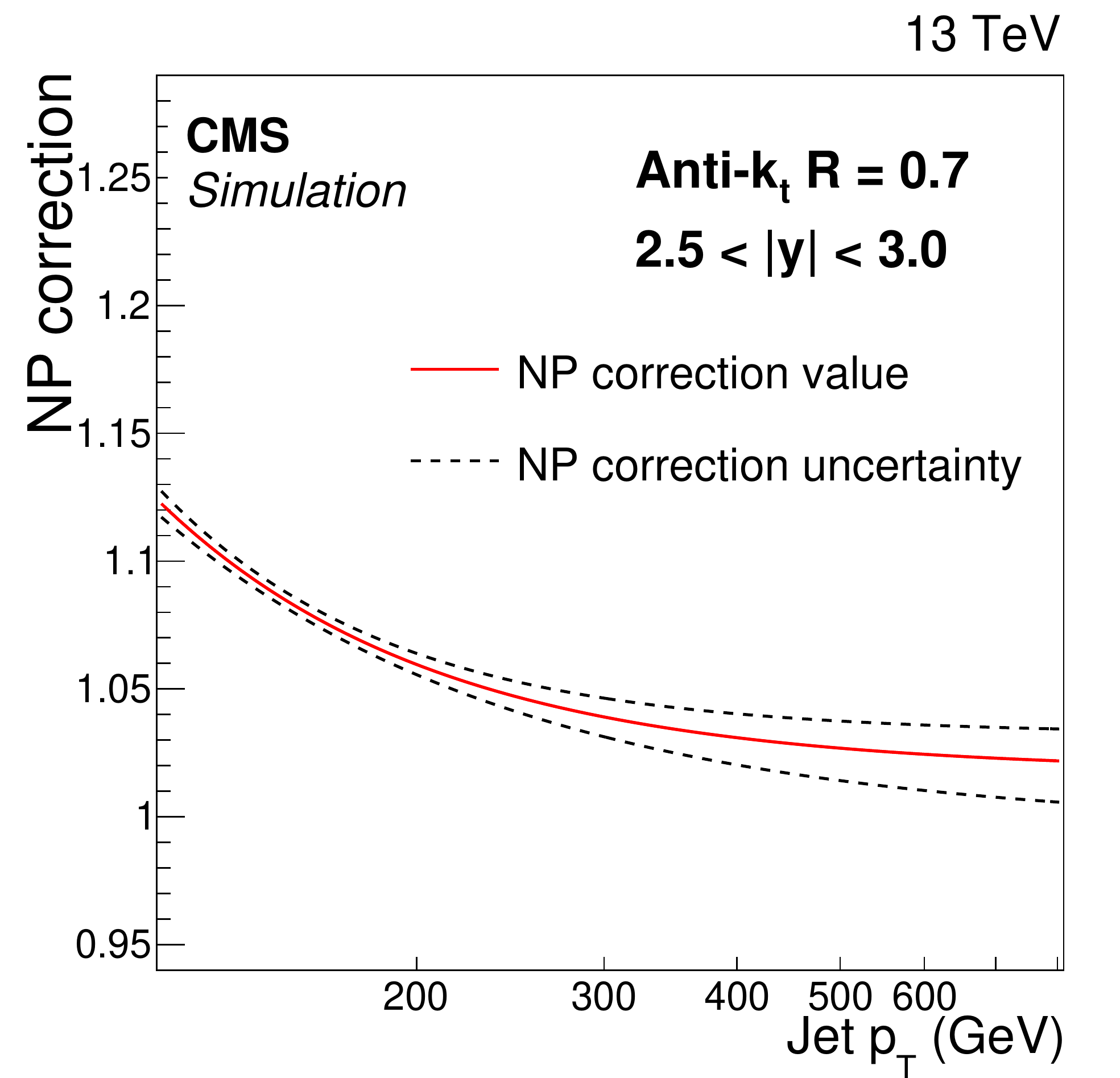}\\
  \caption{Fits to the nonperturbative corrections obtained for
    inclusive AK7 jet cross sections as a function of jet
    $\pt$ for two rapidity bins: $0.5 < \abs{y} < 1.0$
    (left) and $2.5 < \abs{y} < 3.0$ (right). The dotted lines
    represent the uncertainty bands, which are evaluated by fitting the
    envelopes of the predictions of the different generators used.}
  \label{npcorr1}
\end{figure*}

\begin{figure*}[th]
  \centering
  \includegraphics[width=0.45\textwidth]{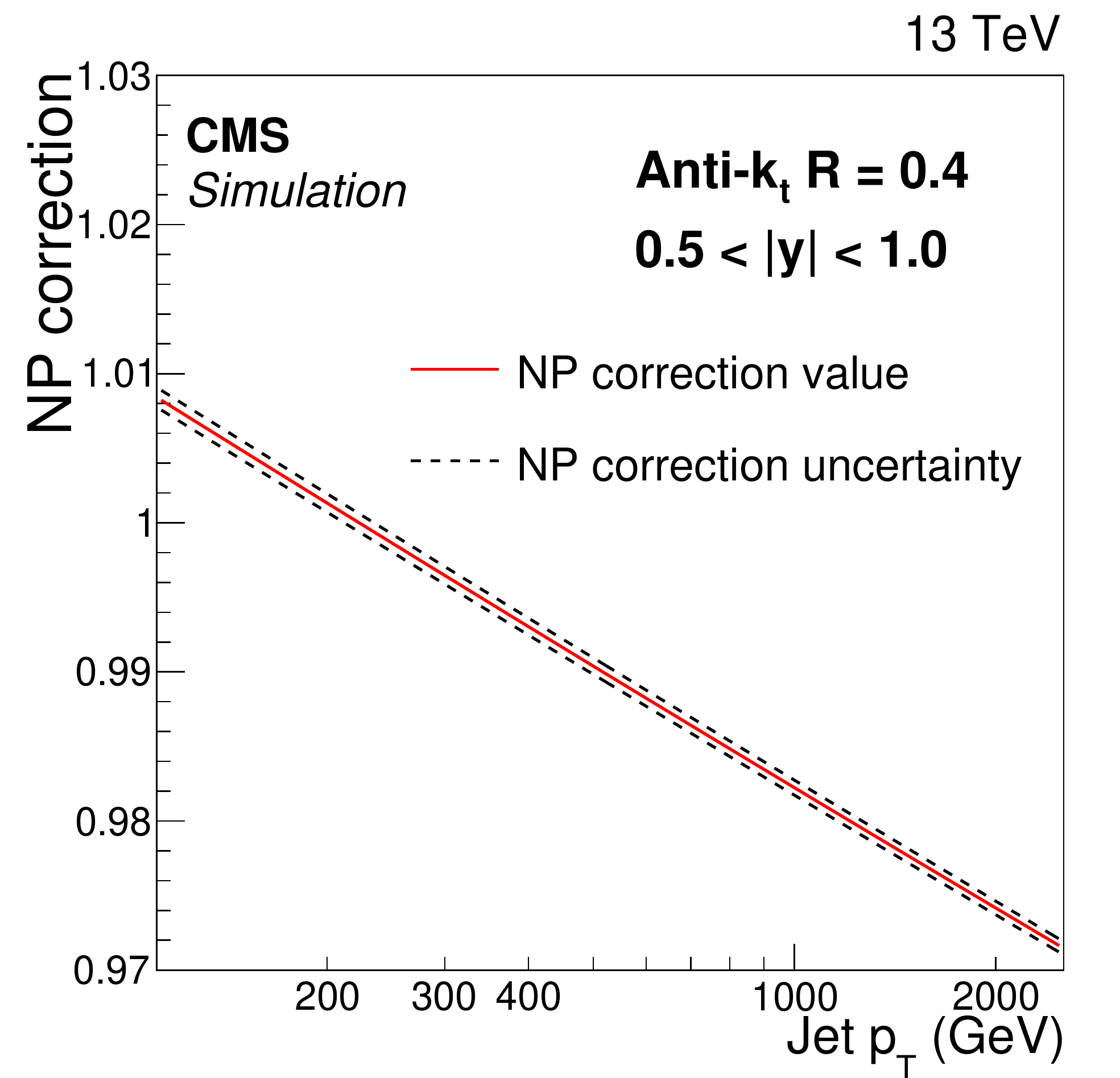}
  \includegraphics[width=0.45\textwidth]{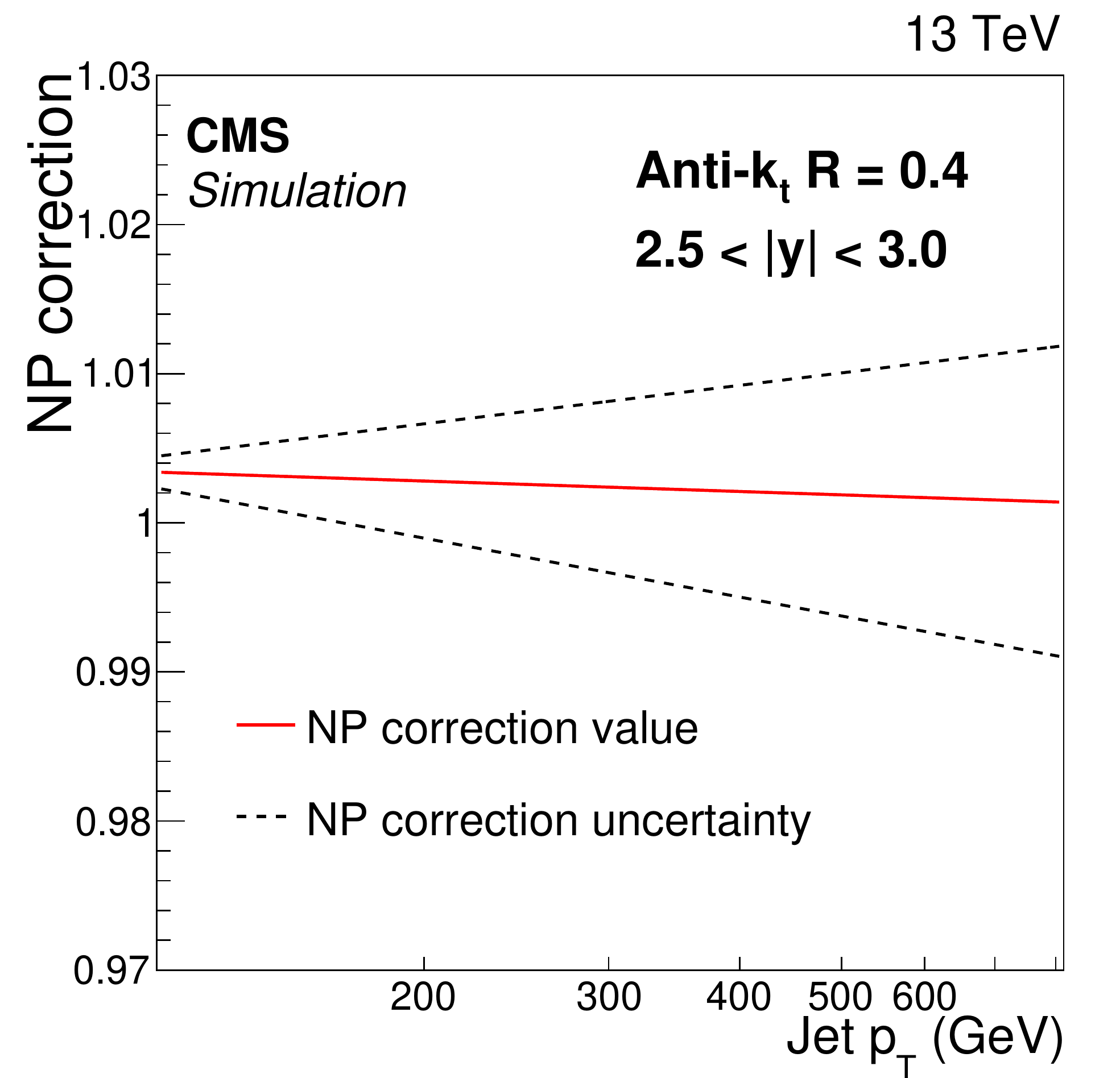}\\
  \caption{Fits to the nonperturbative corrections obtained for
    inclusive AK4 jet cross sections as a function of jet
    $\pt$ for two rapidity bins: $0.5 < \abs{y} < 1.0$
    (left) and $2.5 < \abs{y} < 3.0$ (right). The dotted lines
    represent the uncertainty bands, which are evaluated by fitting the
    envelopes of the predictions of the different generators used.}
  \label{npcorr2}
\end{figure*}

\begin{figure*}[t]
  \centering
  \includegraphics[width=0.45\textwidth]{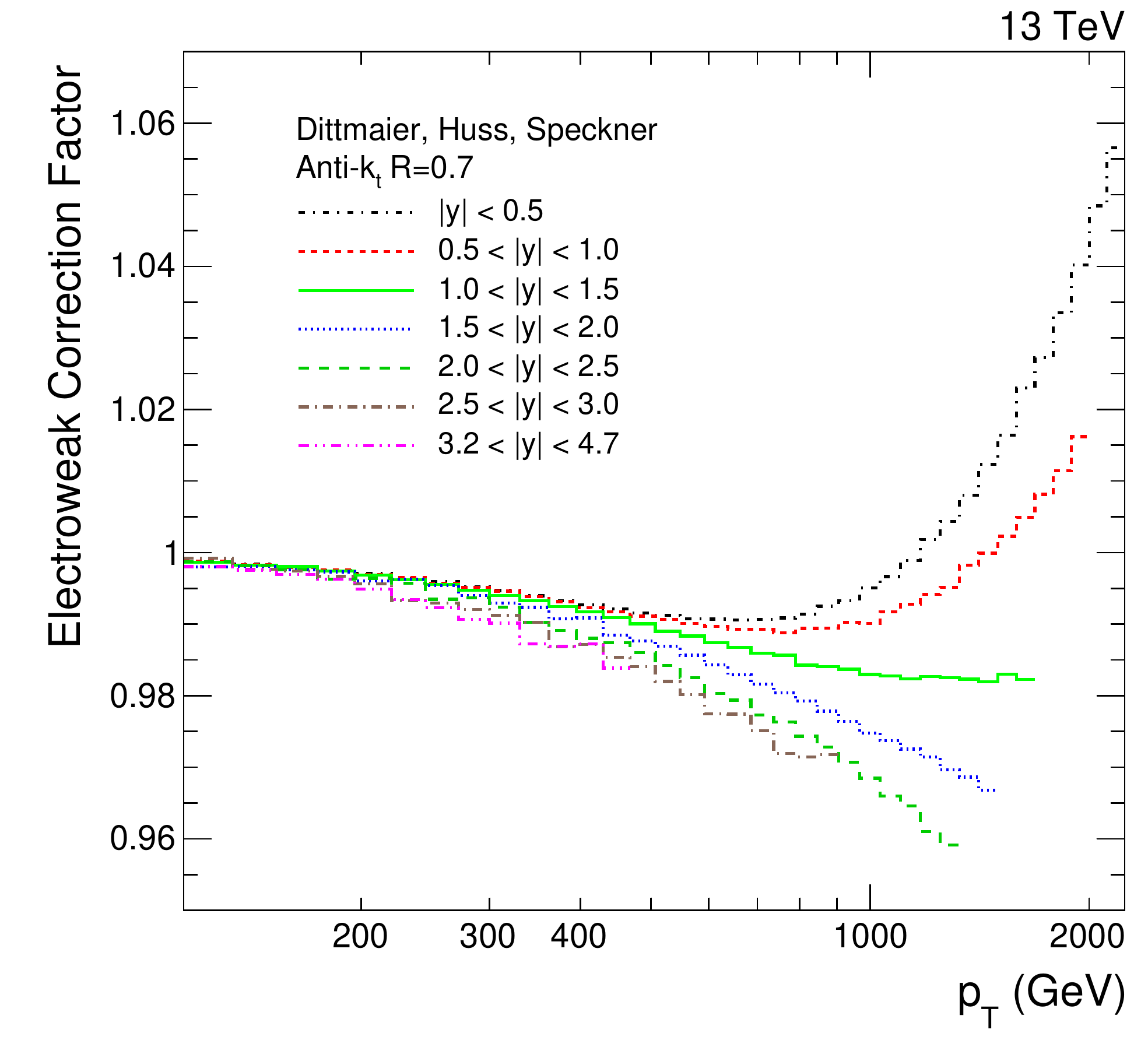}
  \includegraphics[width=0.45\textwidth]{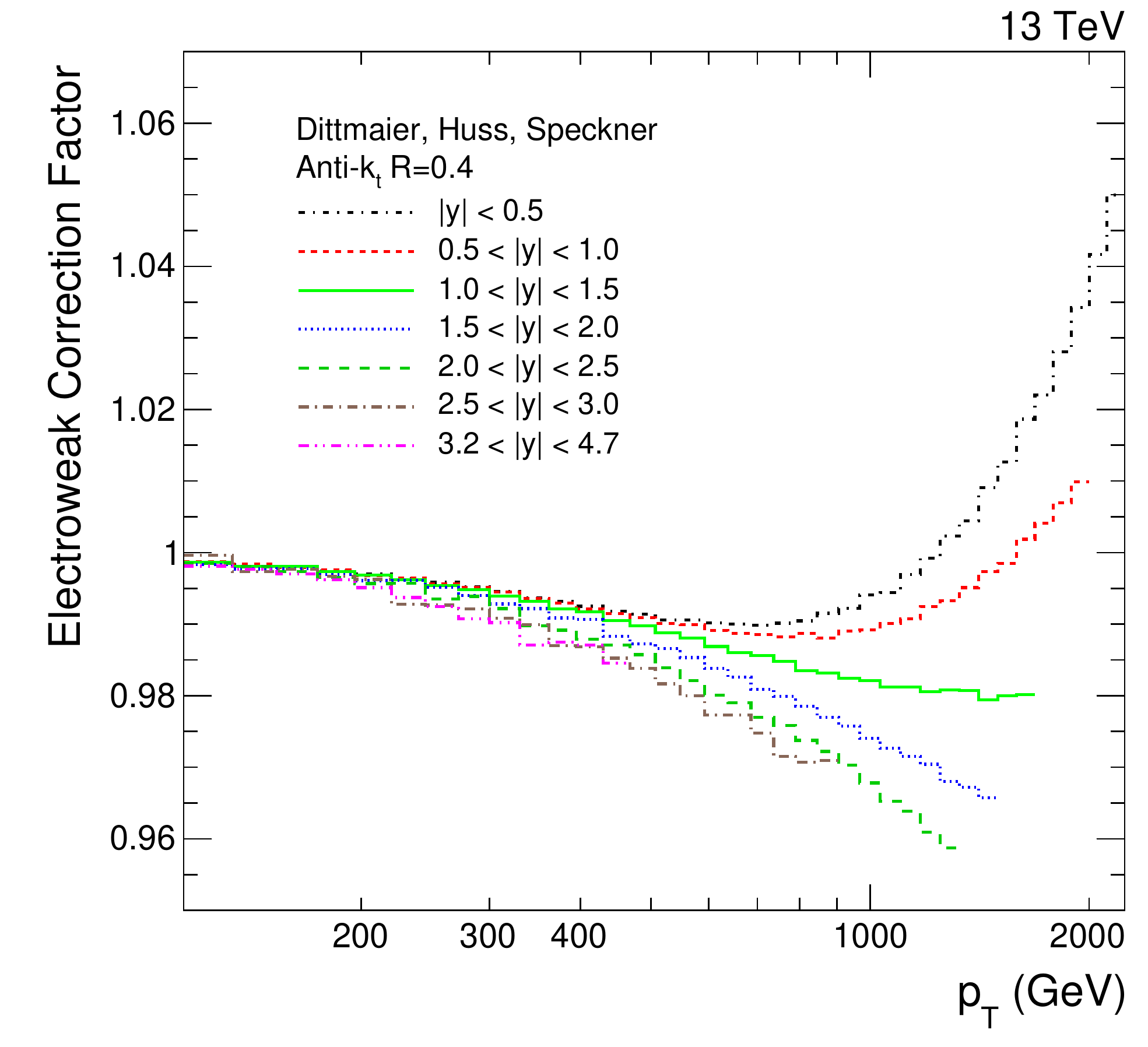}\\
  \caption{Electroweak correction factors for the seven rapidity bins
    for the AK7 (left) and AK4 (right) jets as a function of jet
    $\pt$.}
  \label{ewcorr1}
\end{figure*}

The theoretical predictions for the jet cross section are calculated at NLO
accuracy in pQCD and are evaluated by using \NLOJETPP within the
framework of \fastNLO. The cross sections are calculated at NLO for
single inclusive jet production. The renormalisation and the
factorisation scales ($\mur$ and $\muf$) are chosen to be equal
to the jet $\pt$. Five quarks are assumed to be
massless in the calculation, which is performed using four different
PDF sets with NLO accuracy: CT14~\cite{Dulat:2015mca},
HERAPDF1.5~\cite{Aaron:2009aa}, MMHT2014~\cite{Harland-Lang:2014zoa},
and NNPDF3.0~\cite{Ball:2014uwa}, with the default values of the
strong coupling $\alpsmz=0.1180$, $0.1176$, $0.1200$, and $0.1180$,
respectively.

The theoretical uncertainties are evaluated as the qua\-dra\-tic sum of
the scale, PDF, \alps, and NP uncertainties. The scale uncertainty is
calculated by varying $\mur$ and $\mu_{f}$ in the following six
combinations: (\mur/\pt, \muf/\pt) = (1/2,1/2), (1/2,1),
(1,1/2), (1,2), (2,1) and (2,2). The (asymmetric) scale uncertainty
is determined through the maximal upwards and downwards deviations
with respect to cross sections obtained with the default setting. The
PDF and \alps uncertainties are calculated according to the
prescription of CT14 at the 90\% confidence level and scaled down to a
68.3\% confidence level.

The impact of NP effects, \ie MPI and HAD effects, is evaluated by
using samples obtained from different MC event generators with a
simulation of PS and MPI contributions. The following MC event
generators are used to estimate the NP corrections: LO
\PYTHIAE with tune CUETM1, LO \HERWIGpp 2.7.0~\cite{Bellm:2013hwb}
with tunes UE-EE-5C~\cite{Seymour:2013qka} and
CUETS1~\cite{Khachatryan:2015pea}, and NLO
\POWHEG~\cite{Nason:2004rx,Frixione:2007vw,Alioli:2010xa}. The matrix
element calculation performed with \POWHEG is interfaced to \PYTHIAE
with three different tunes (CUETS1-CTEQ6L1, CUETS1-HERAPDF, and
CUETM1) for the simulation of the underlying-event (UE) contributions.
The cross section ratios between a nominal event generation interfaced
to the simulation of UE contributions, and a sample without HAD and
MPI effects are taken as correction separately in each considered
rapidity range. In a compact formulation, the NP correction factors
can be defined as
\begin{equation}\label{old}
  C^{\mathrm{\tiny{NP}}}= \frac{   \rd \sigma^{\mathrm{\tiny{PS+HAD+MPI}}} / \rd \pt } { \rd \sigma^{\mathrm{\tiny{PS}}} / \rd \pt},
\end{equation}
where $\sigma^{\mathrm{\tiny{PS+HAD+MPI}}}$ is the cross section
obtained with an MC sample simulating the contribution of PS, HAD, and
MPI, while $\sigma^{\mathrm{\tiny{PS}}}$ includes only PS
effects. Corrections obtained with various NLO and LO event generators
are evaluated separately for the AK7 and AK4 jets. The average of the
results from the NLO and LO event generators defines the central value
of the NP corrections, which are fitted to a power-law function in jet
\pt. The uncertainty in the NP corrections are evaluated by
fitting the upper and lower values of the predictions of the different
generators. The combinations of PDF sets, matrix element calculations,
and UE tunes used to evaluate the NP corrections are validated on UE,
minimum bias and jet variables, and they are able to reproduce a
wide set of observables~\cite{Khachatryan:2015pea}. The NP corrections
are shown in Figs.~\ref{npcorr1} and~\ref{npcorr2}, respectively, for
AK7 and AK4 jets in a central ($0.5 < \abs{y} < 1.0$) and a forward
rapidity bin ($2.5 < \abs{y} < 3.0$).

The NP corrections for the AK7 jets are $\approx$ 15\% (13\%) for
\pt $\sim$~114\GeV in the region $0.5<\abs{y}<1.0$ ($2.5<\abs{y}<3.0$) and decrease rapidly for increasing \pt, flattening at
values of~${\approx}1$ for $\pt\sim200$--300\GeV, depending on
the considered rapidity range. Because of the smaller cone size, AK4
jets are less affected by the MPI and HAD effects. In particular, the
additional energy produced by MPI shrinks for decreasing radii R,
while the out-of-cone losses due to HAD effects increase for smaller
radii R. These two effects are responsible for NP corrections that
fall below 1 for AK4 jets with $\pt >200\GeV$ at central
rapidity. The NP corrections for AK4 jets are very close to unity in the
phase space considered. For both cone sizes, the uncertainty assigned
to the NP corrections is of the order of 1--2\%.

\begin{figure*}[htbp]
  \includegraphics[width=0.5\textwidth]{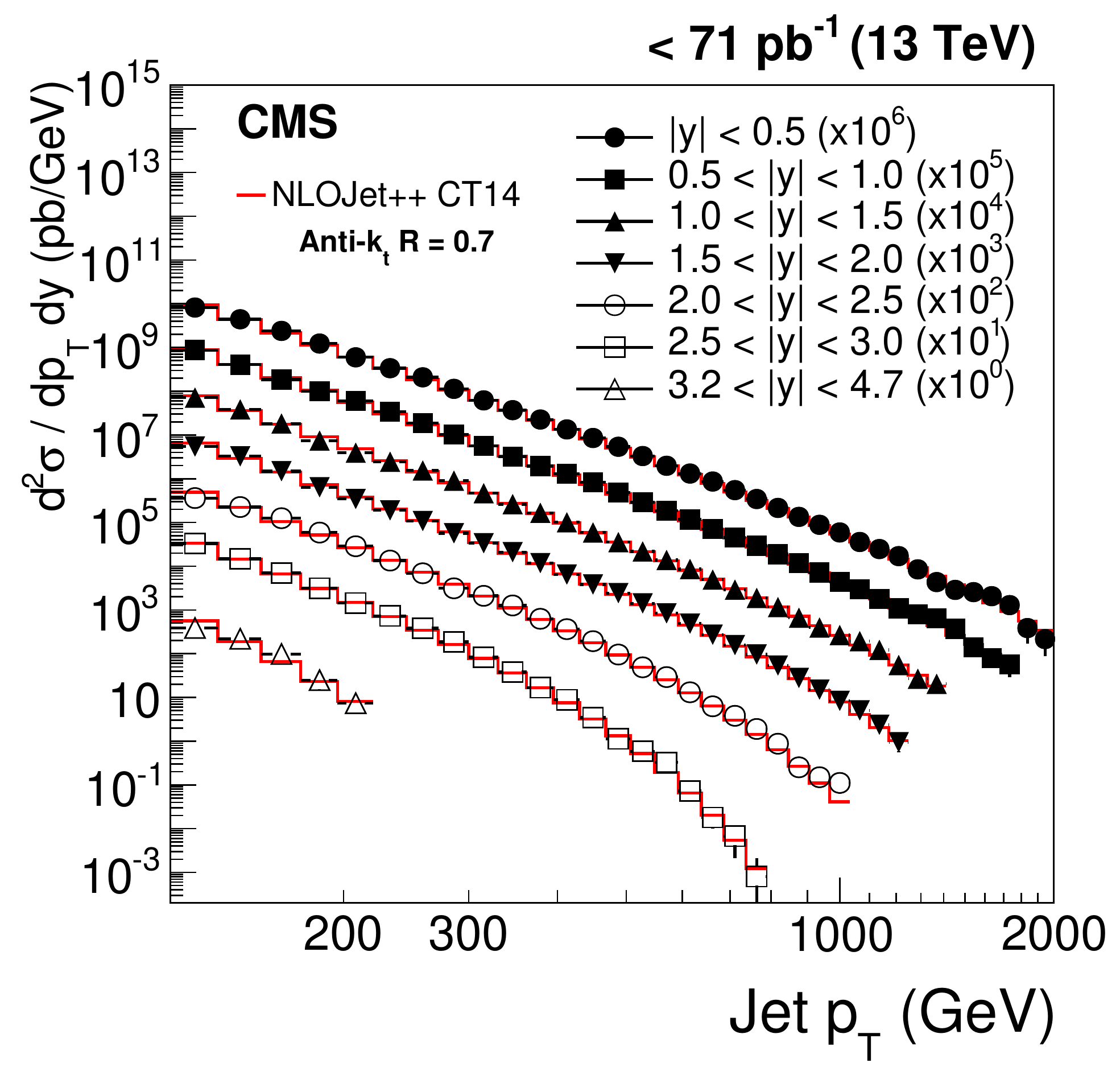}
  \includegraphics[width=0.5\textwidth]{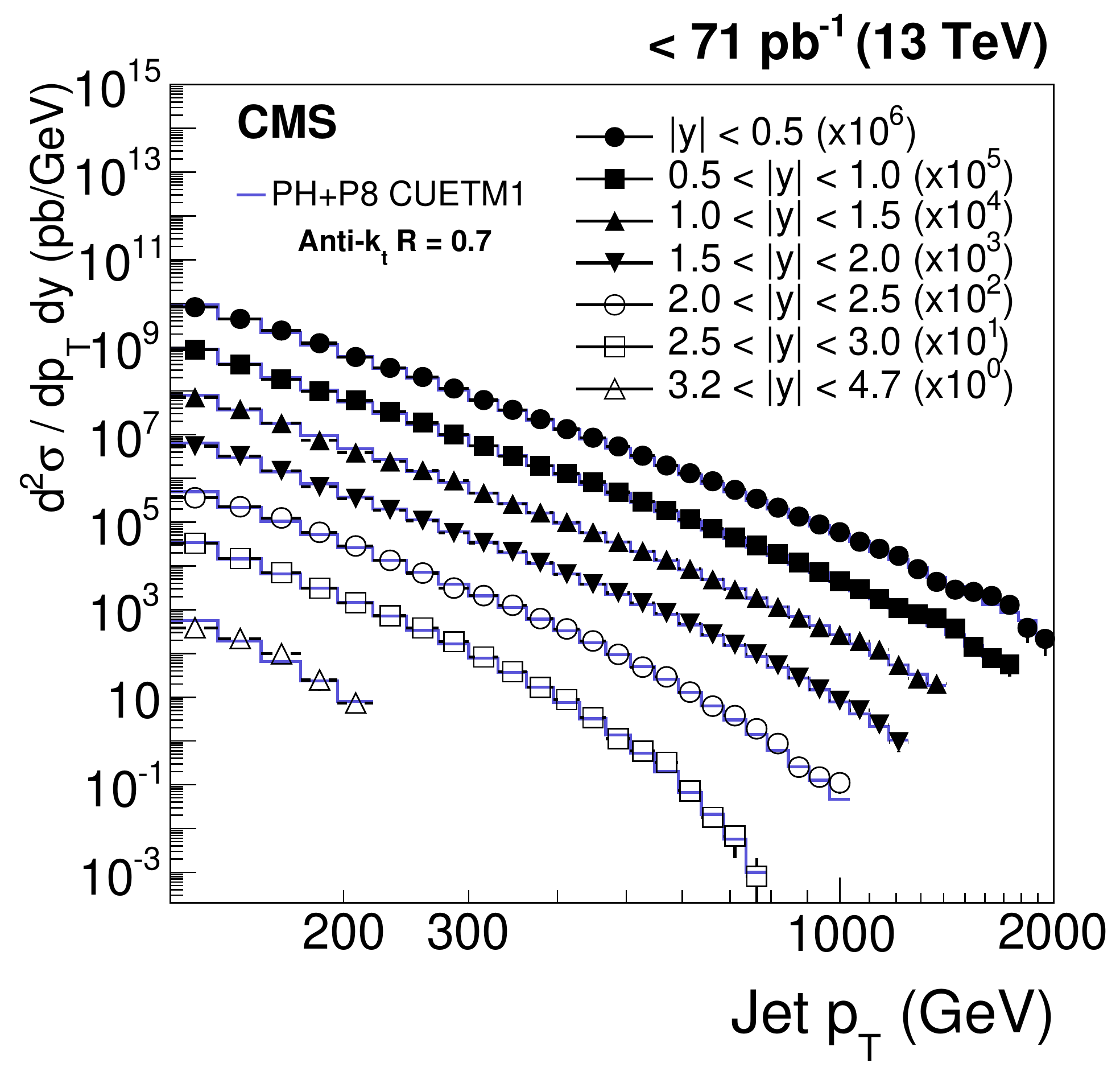}
  \caption{Double-differential inclusive jet cross section as function
    of jet \pt. On the left, data (points) and predictions from
    \NLOJETPP based on the CT14 PDF set corrected for the NP and
    electroweak effects (line) are shown. On the right, data (points)
    and predictions from \POWHEG (PH) + \PYTHIAE (P8) with tune CUETM1
    (line) are shown. Jets are clustered with the anti-\kts algorithm
    ($R = 0.7$).}
  \label{fig:crossSectionAK7}
\end{figure*}

\begin{figure*}[htbp]
  \includegraphics[width=0.5\textwidth]{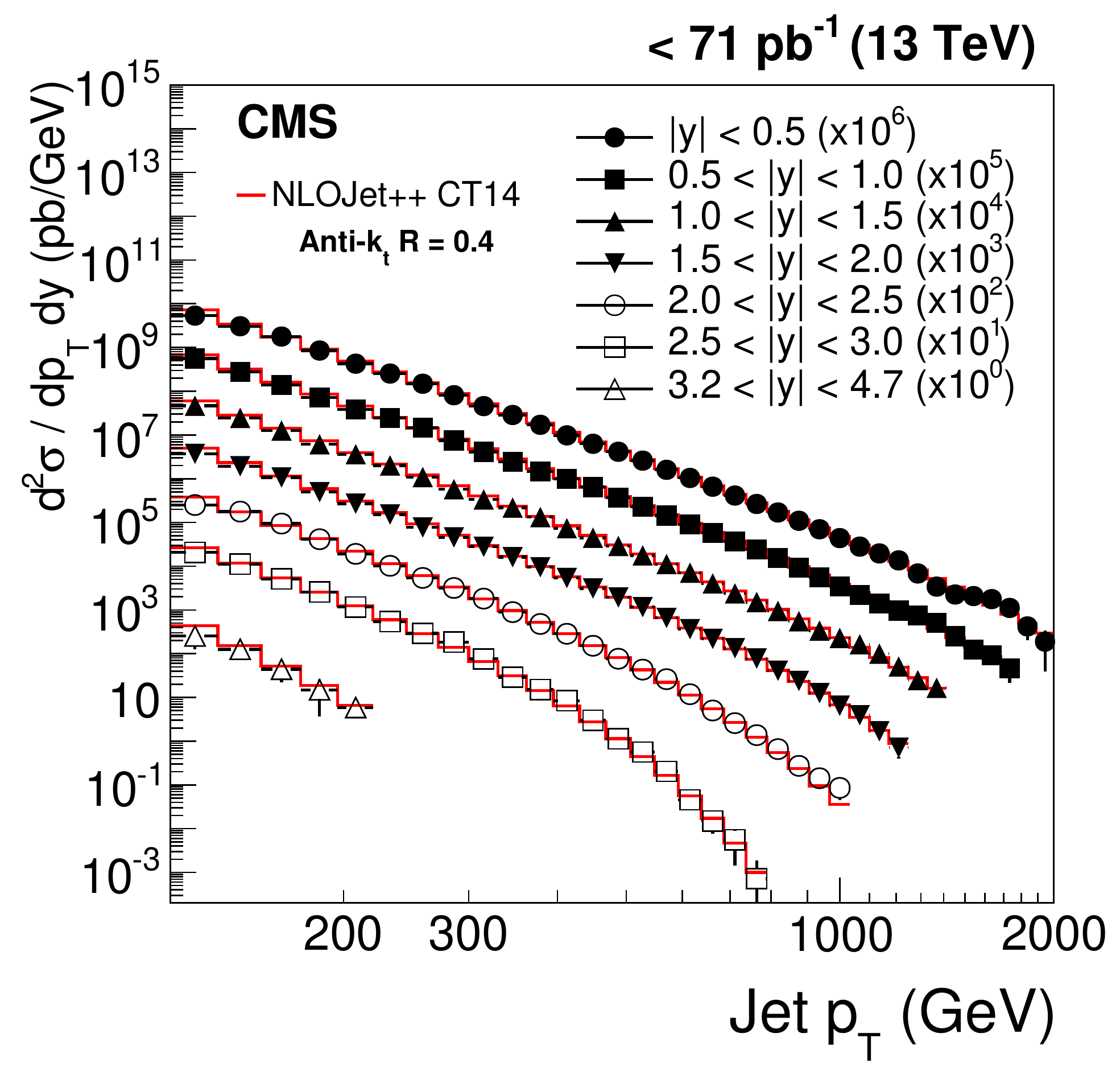}
  \includegraphics[width=0.5\textwidth]{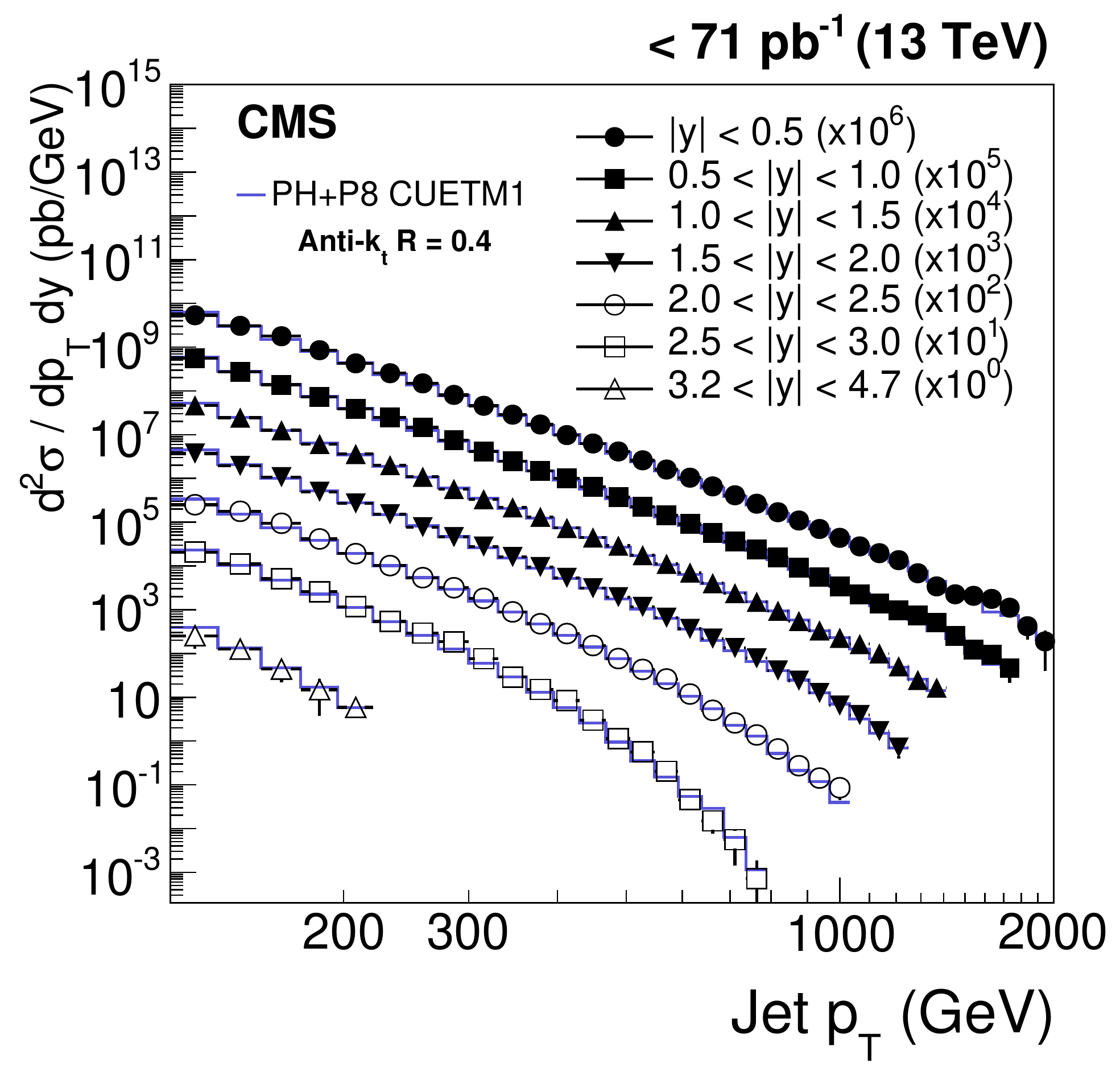}
  \caption{Double-differential inclusive jet cross section as function
    of jet \pt. On the left, data (points) and predictions from
    \NLOJETPP based on the CT14 PDF set corrected for the NP and
    electroweak effects (line) are shown.  On the right, data (points)
    and predictions from \POWHEG (PH) + \PYTHIAE (P8) with tune CUETM1
    (line) are shown.  Jets are clustered with the anti-\kts algorithm
    ($R = 0.4$).}
  \label{fig:crossSectionAK4}
\end{figure*}

Electroweak effects, which arise from the virtual exchanges of massive
gauge W and Z bosons, become sizable at high jet \pt and
central rapidity. Corrections to electroweak effects are shown in
Fig.~\ref{ewcorr1} for both AK7 and AK4
jets~\cite{Dittmaier:2012kx}. They range between 0.96 and 1.05,
depending on the jet \pt and rapidity, and are less than 3\% for
$\pt <1\TeV$ and very similar between the two cone sizes. For
jet measurements performed at a centre-of-mass energy of
7\TeV~\cite{Khachatryan:2014waa},
electroweak corrections of 10--15\% are observed for jet
$\pt >1\TeV$ in the $\abs{y}<1.0$ range, decreasing below 2\%
for lower \pt, independent of the jet rapidity.
Electroweak corrections are applied to the \NLOJETPP predictions in a similar manner to the NP contributions.

\subsection{Predictions from fixed-order calculations matched to
  parton shower simulations}
\label{sec:compPowheg}

The predictions from different MC event generators are compared to
data. The \HERWIGpp and the \PYTHIAE event generators are
considered. Both of them are based on an LO $2\to2$ matrix
element calculation. The \PYTHIAE event generator simulates parton showers ordered in
\pt and uses the Lund string model~\cite{Andersson:1998tv} for
HAD, while \HERWIGpp generates parton showers through angular-ordered
emissions and uses a cluster fragmentation model \cite{Webber:1983if}
for HAD\@.  The contribution of MPI is simulated in both \PYTHIAE and
\HERWIGpp\@. In particular, \PYTHIAE applies a model \cite{Corke:2010yf}
where MPI are interleaved with parton showering, while \HERWIGpp
models the overlap between the colliding protons through a Fourier
transform of the electromagnetic form factor, which plays the role of
an effective inverse proton radius. Depending on the amount of proton
overlap, the contribution of generated MPI varies in the simulation.
The MPI parameters of both generators are tuned to measurements in
proton-proton collisions at the LHC~\cite{Khachatryan:2015pea}, while
the HAD parameters are determined from fits to LEP data. For \PYTHIAE,
the CUETM1 tune, which is based on NNPDF2.3LO~\cite{Ball:2013hta,Ball:2011uy}, is
considered, while \HERWIGpp uses the CUETS1 tune~\cite{Khachatryan:2015pea}, based on the CTEQ6L1
PDF set~\cite{Pumplin:2002vw}.

Predictions based on NLO pQCD are also considered using the \POWHEG
package
matched to \PYTHIAE\ parton showers and including a simulation of
MPI\@. The \POWHEG sample uses the CT10nlo PDF set~\cite{Lai:2010vv}.
Various tunes in \PYTHIAE are used for the UE simulation, which differ
in the choice of the PDF set and the HAD parameters: the CUETM1, and
tunes CUETS1-CTEQL1 and CUETS1-HERAPDF, which use the
CTEQ6L1
and the HERAPDF1.5LO~\cite{Sarkar:2014zua} PDF sets, respectively.
The HAD parameters for the CUETM1 tune are taken from the Monash
tune~\cite{Skands:2014pea}, while the 4C tune provides these in both
CUETS1 tunes. All these combinations of \POWHEG matrix element and
UE-simulation tunes reproduce with very high precision the UE and jet
observables at various collision energies~\cite{Khachatryan:2015pea}.

\section{Comparison of theoretical predictions and data}
\label{sec:compTheoData}

Figures~\ref{fig:crossSectionAK7} and~\ref{fig:crossSectionAK4} show
the double-differential inclusive jet cross section measurements,
presented as a function of \pt for seven $\abs{y}$ ranges, after
unfolding for detector effects, using the anti-\kts algorithm with $R =
0.7$ and $0.4$, respectively. The measurements are compared to the
\NLOJETPP predictions based on the CT14 PDF set, corrected for NP and
electroweak effects (left), and to the predictions from \POWHEG +
\PYTHIAE with tune CUETM1 (right). The data are consistent with the
predictions over a wide range of jet \pt from 114\GeV up to
2\TeV.

\begin{figure*}[htbp]
  \centering
  \includegraphics[width=0.48\textwidth]{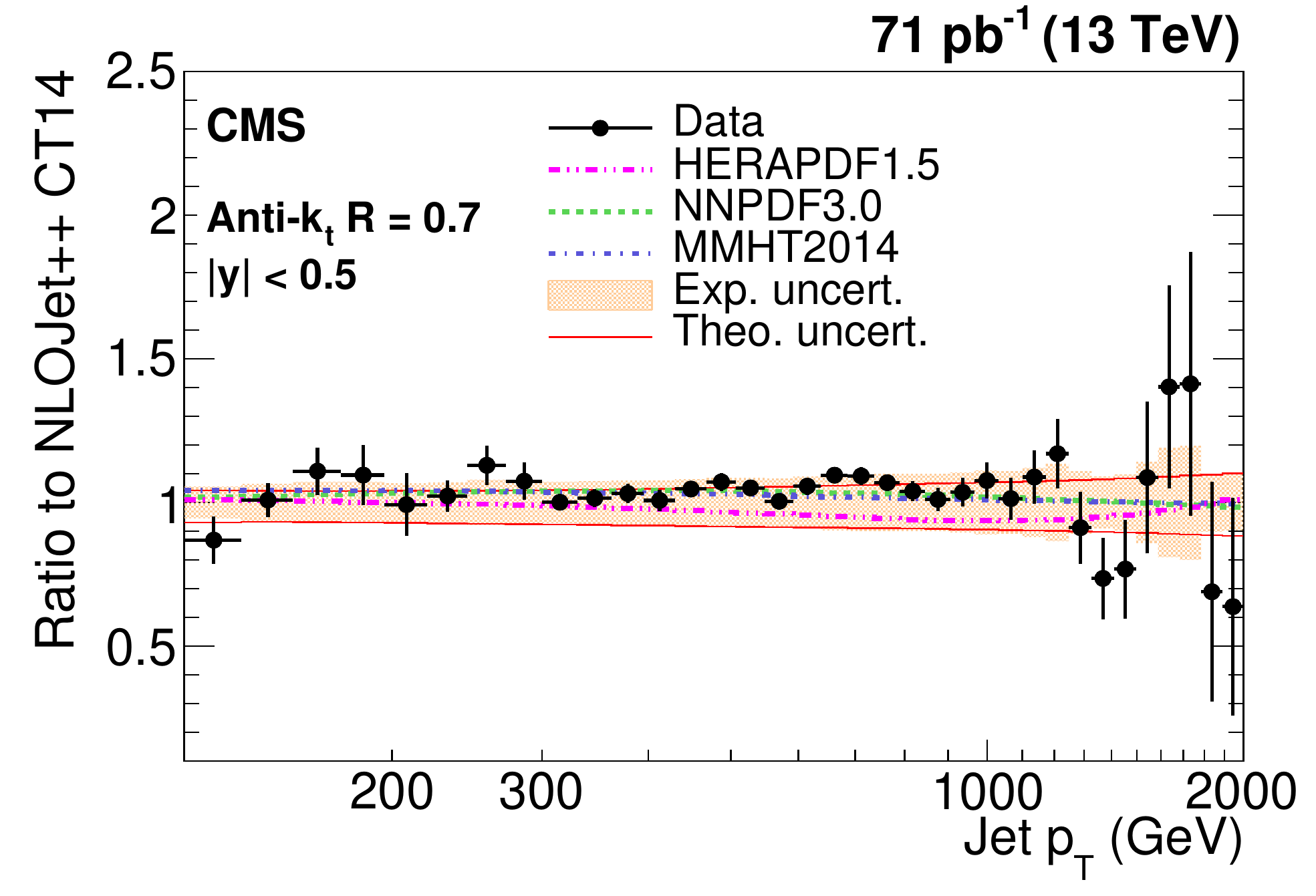}
  \includegraphics[width=0.48\textwidth]{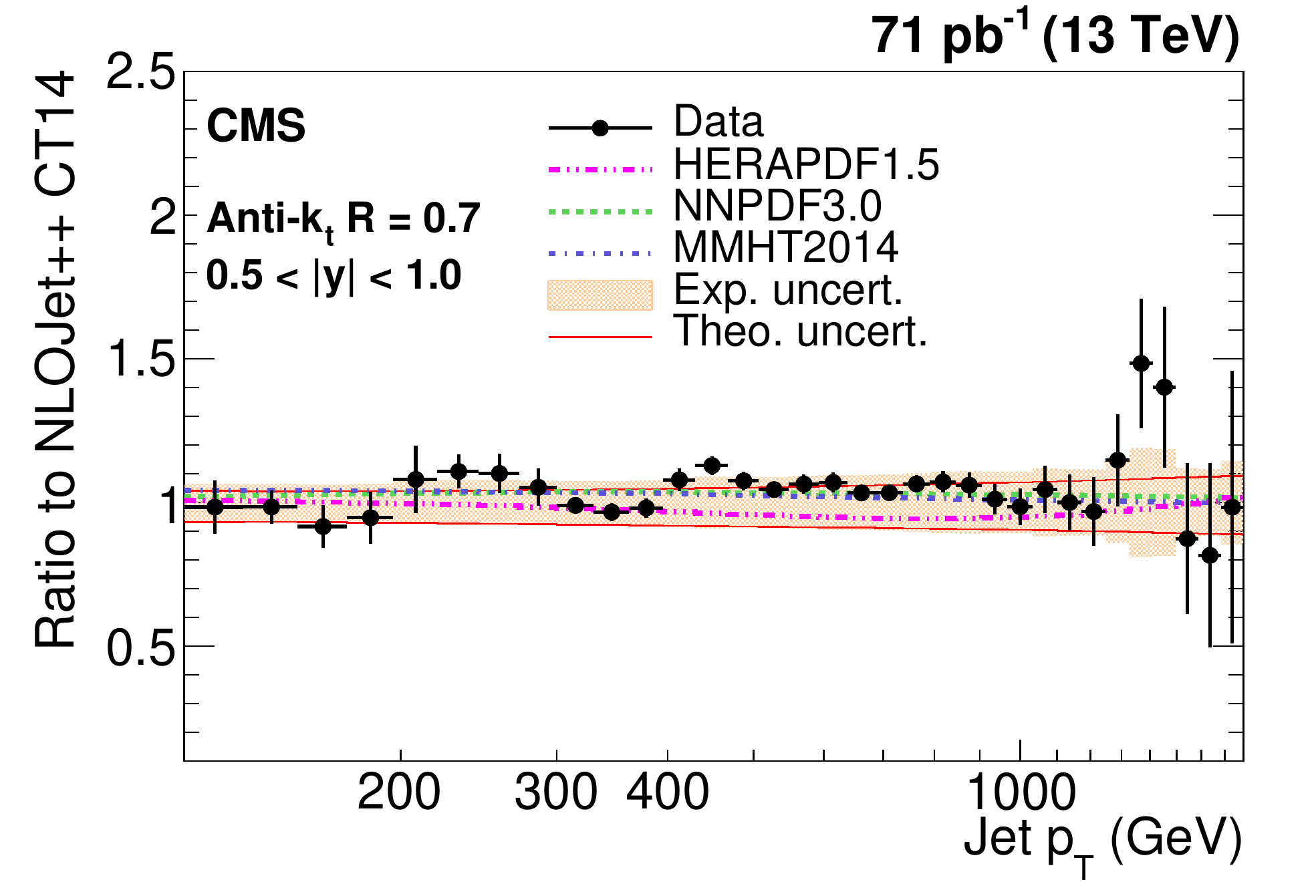}
  \includegraphics[width=0.48\textwidth]{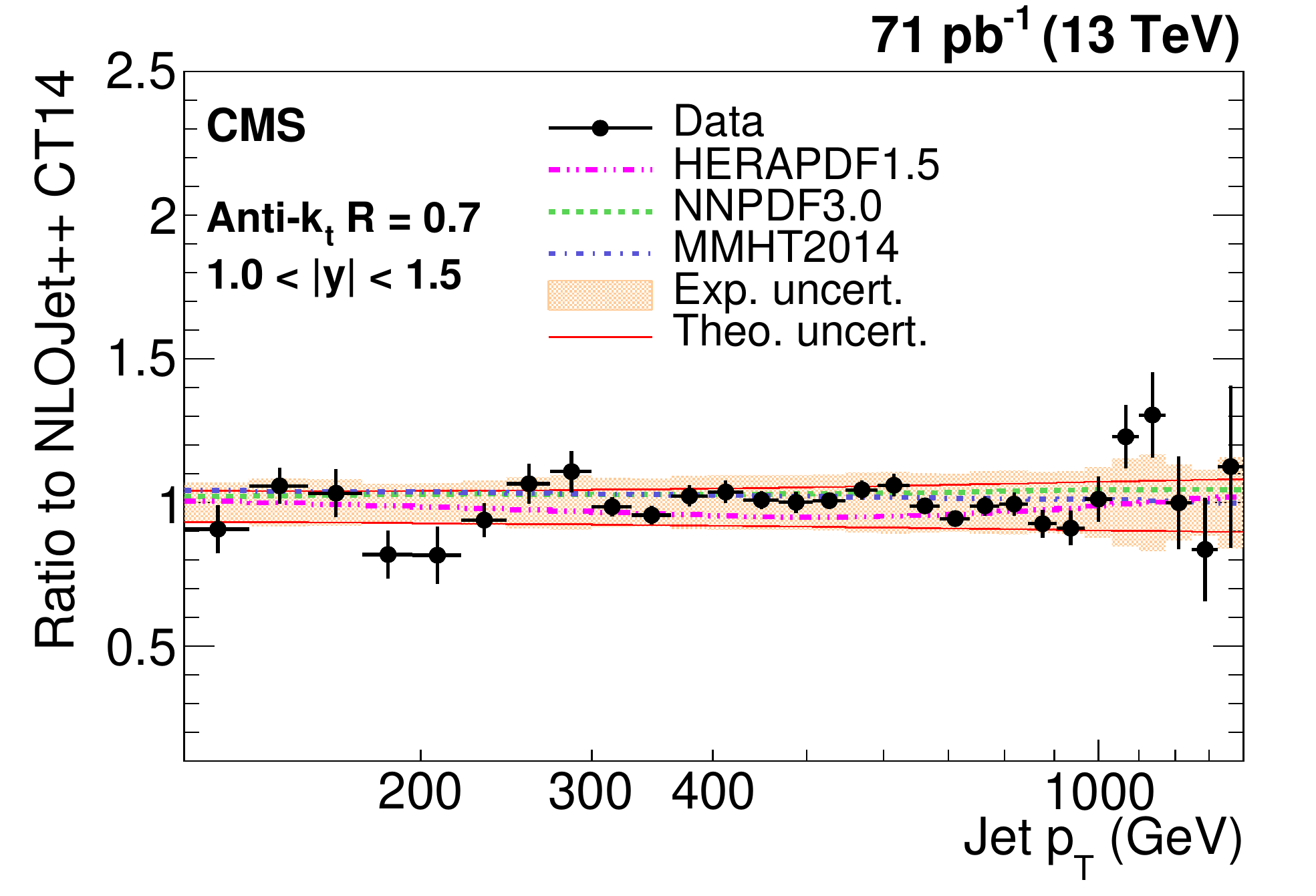}
  \includegraphics[width=0.48\textwidth]{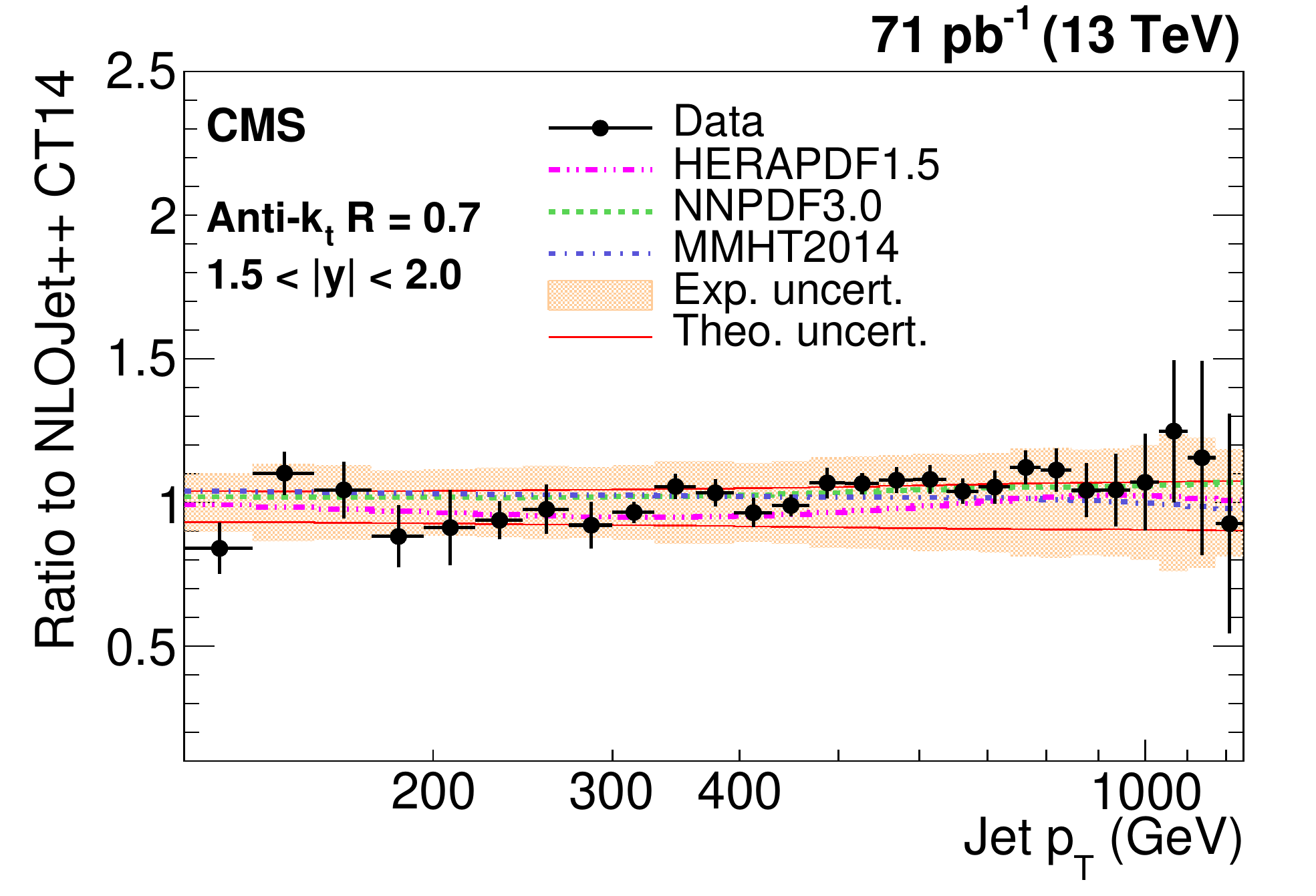}
  \includegraphics[width=0.48\textwidth]{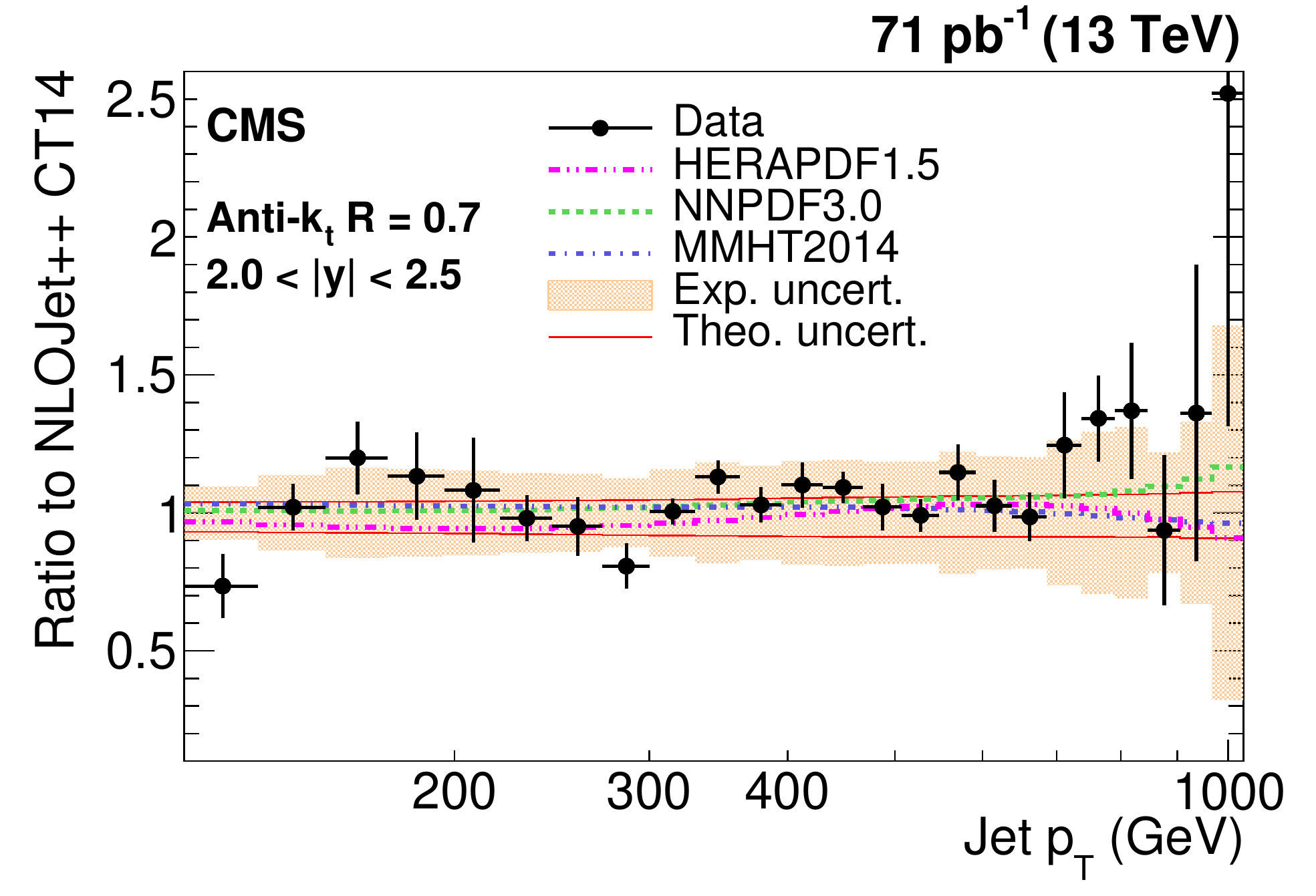}
  \includegraphics[width=0.48\textwidth]{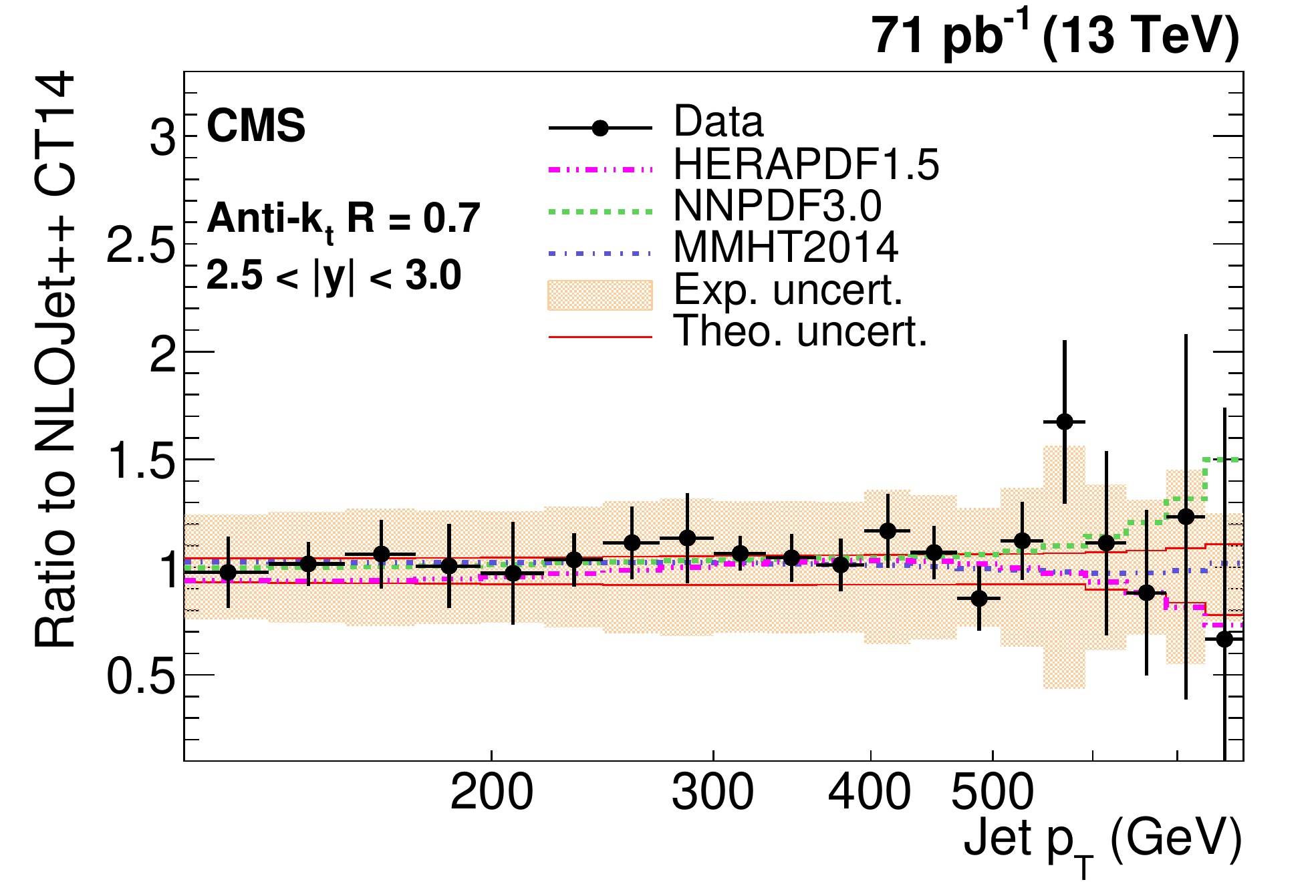}
  \includegraphics[width=0.48\textwidth]{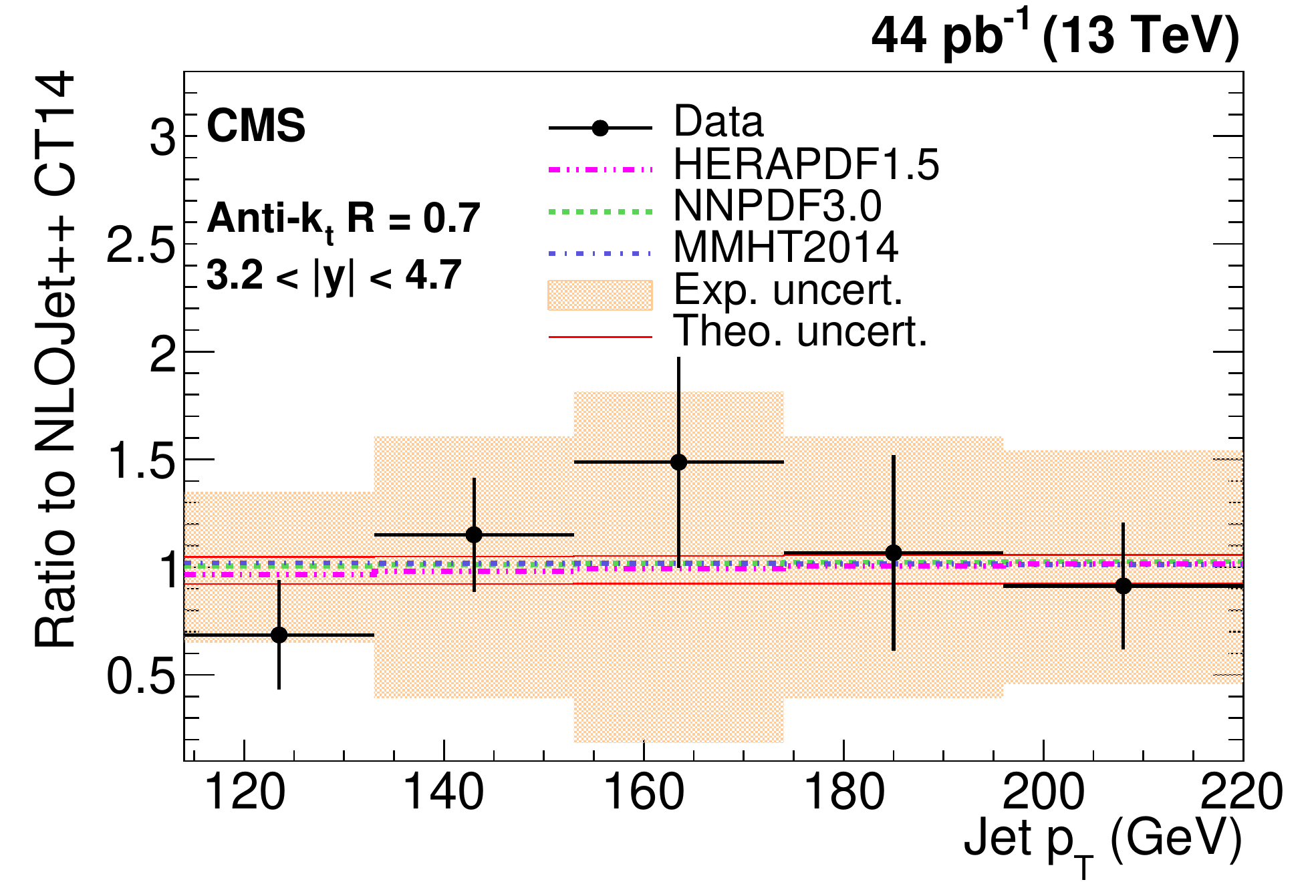}
  \caption{Ratio of measured values to theoretical prediction from
    \NLOJETPP using the CT14 PDF set and corrected for the NP and
    electroweak effects. Predictions employing three other PDF sets
    are also shown for comparison. Jets are clustered with the
    anti-\kts algorithm with a distance parameter of 0.7. The error
    bars correspond to the statistical uncertainties of the data and
    the shaded bands to the total experimental systematic
    uncertainties.}
  \label{fig:AK7_ratio_CT14}
\end{figure*}

\begin{figure*}[htbp]
  \centering
  \includegraphics[width=0.48\textwidth]{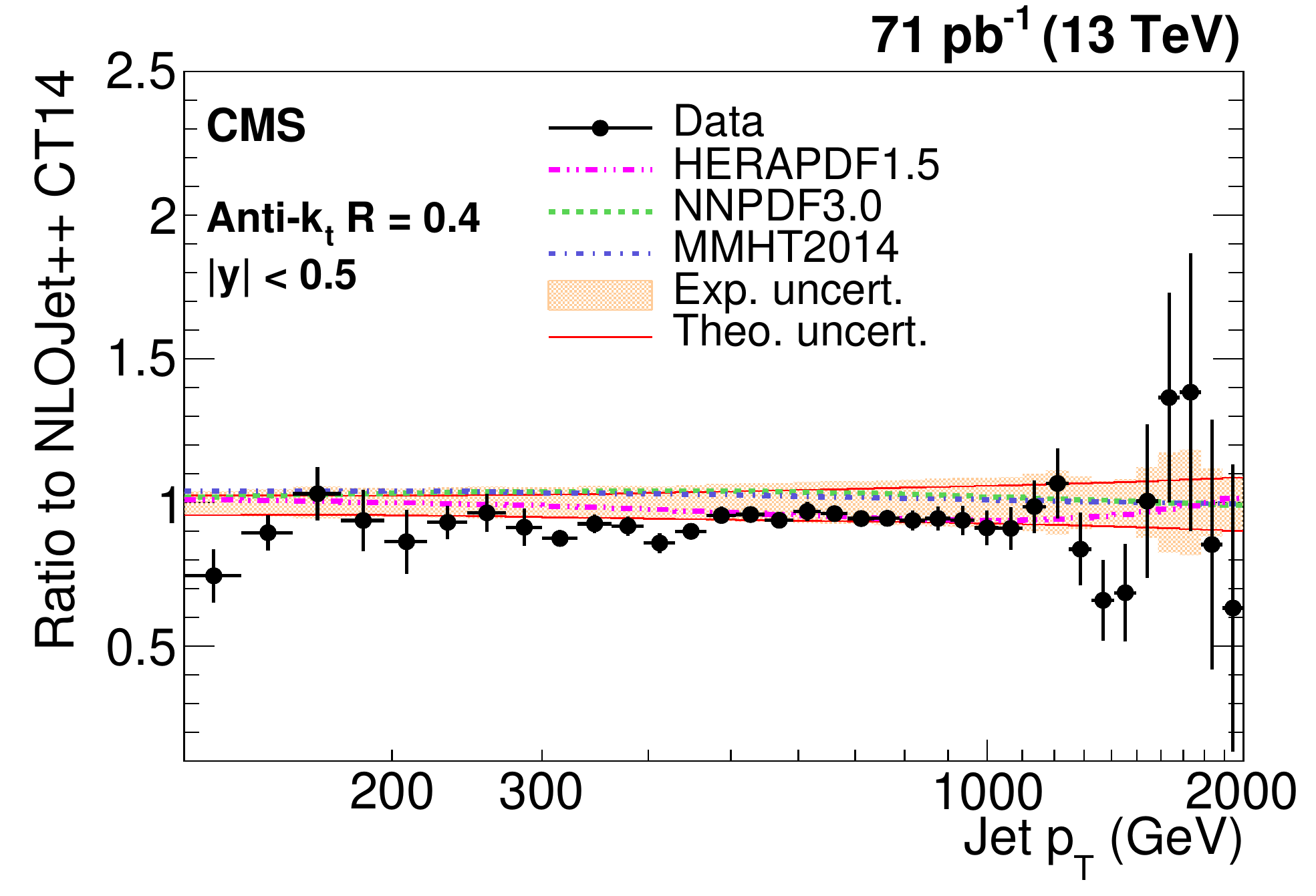}
  \includegraphics[width=0.48\textwidth]{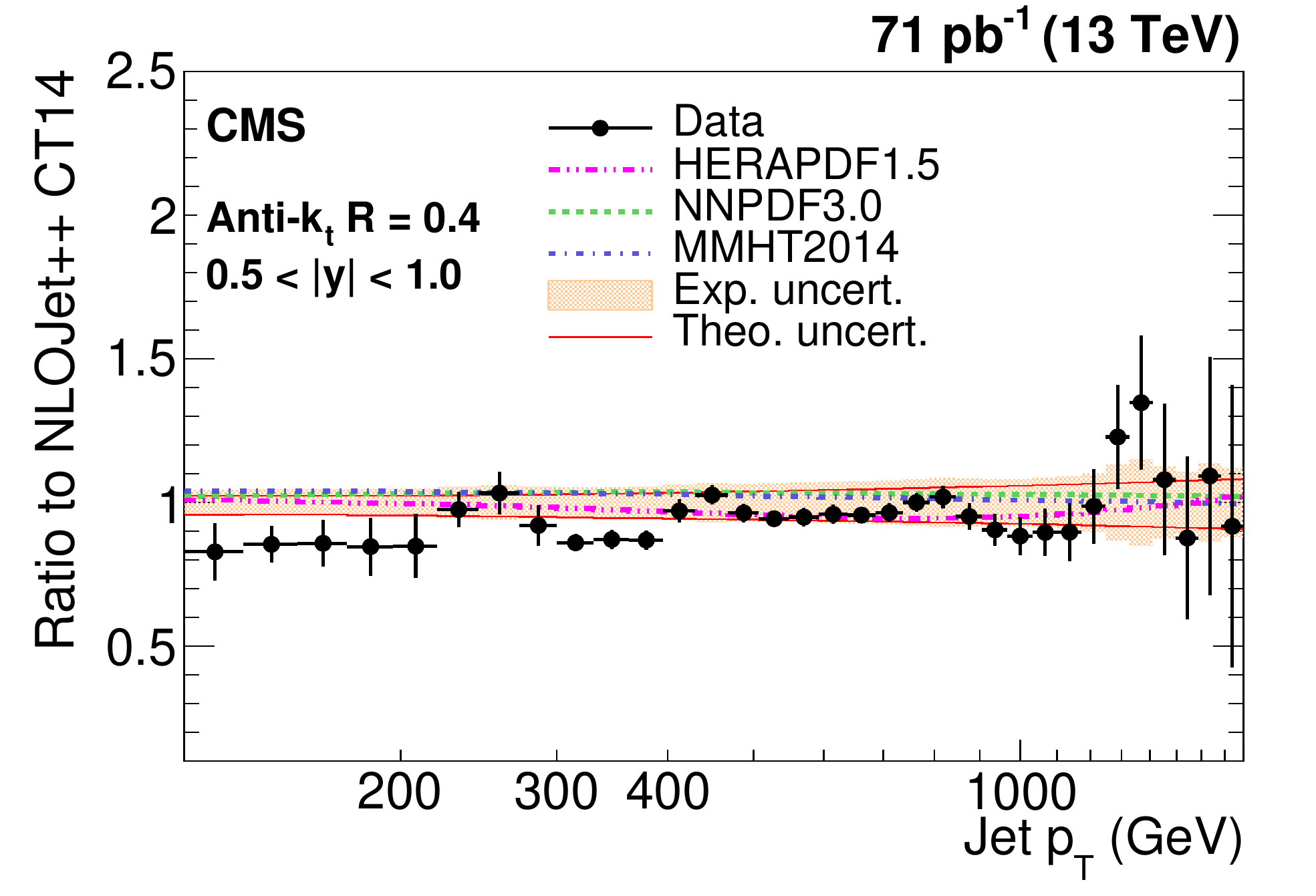}
  \includegraphics[width=0.48\textwidth]{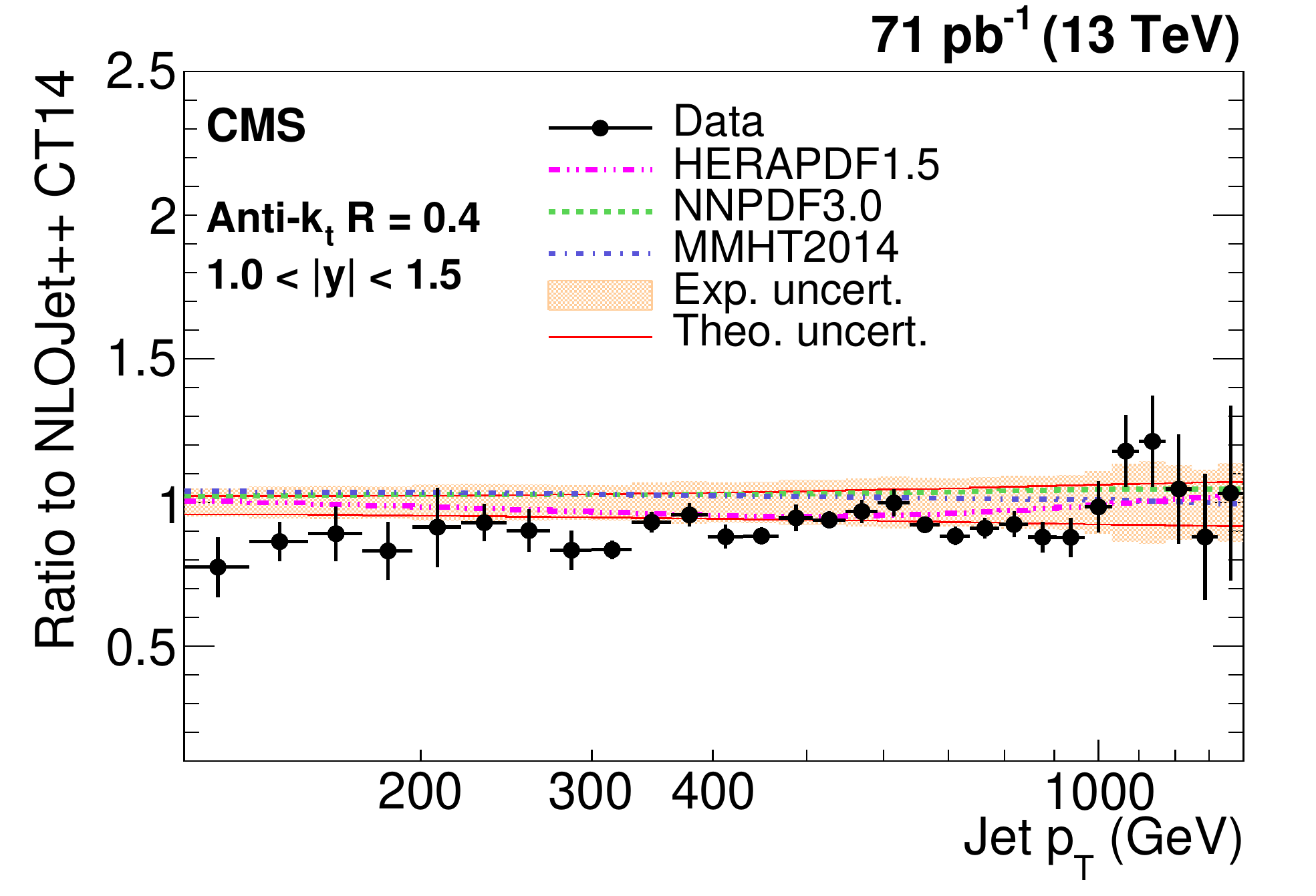}
  \includegraphics[width=0.48\textwidth]{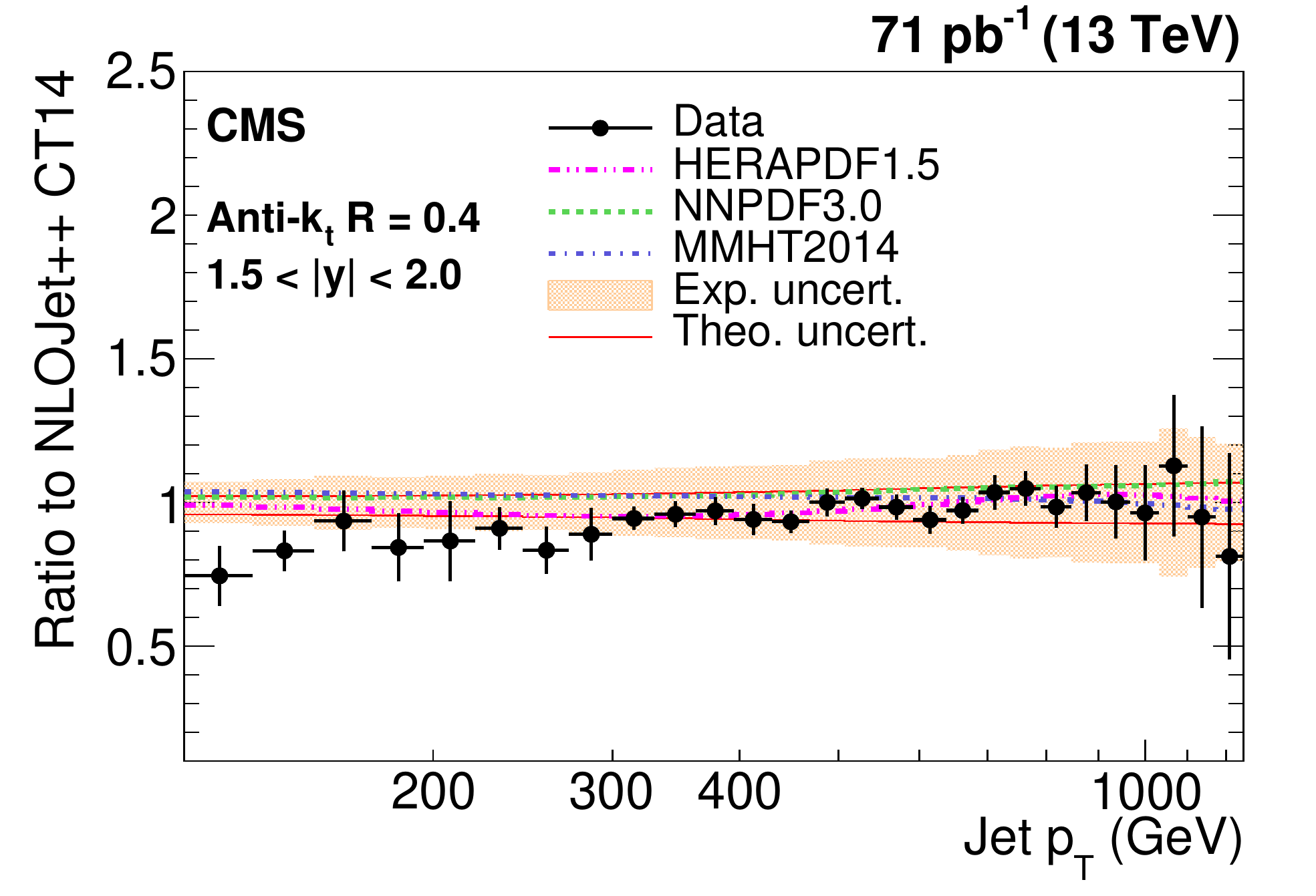}
  \includegraphics[width=0.48\textwidth]{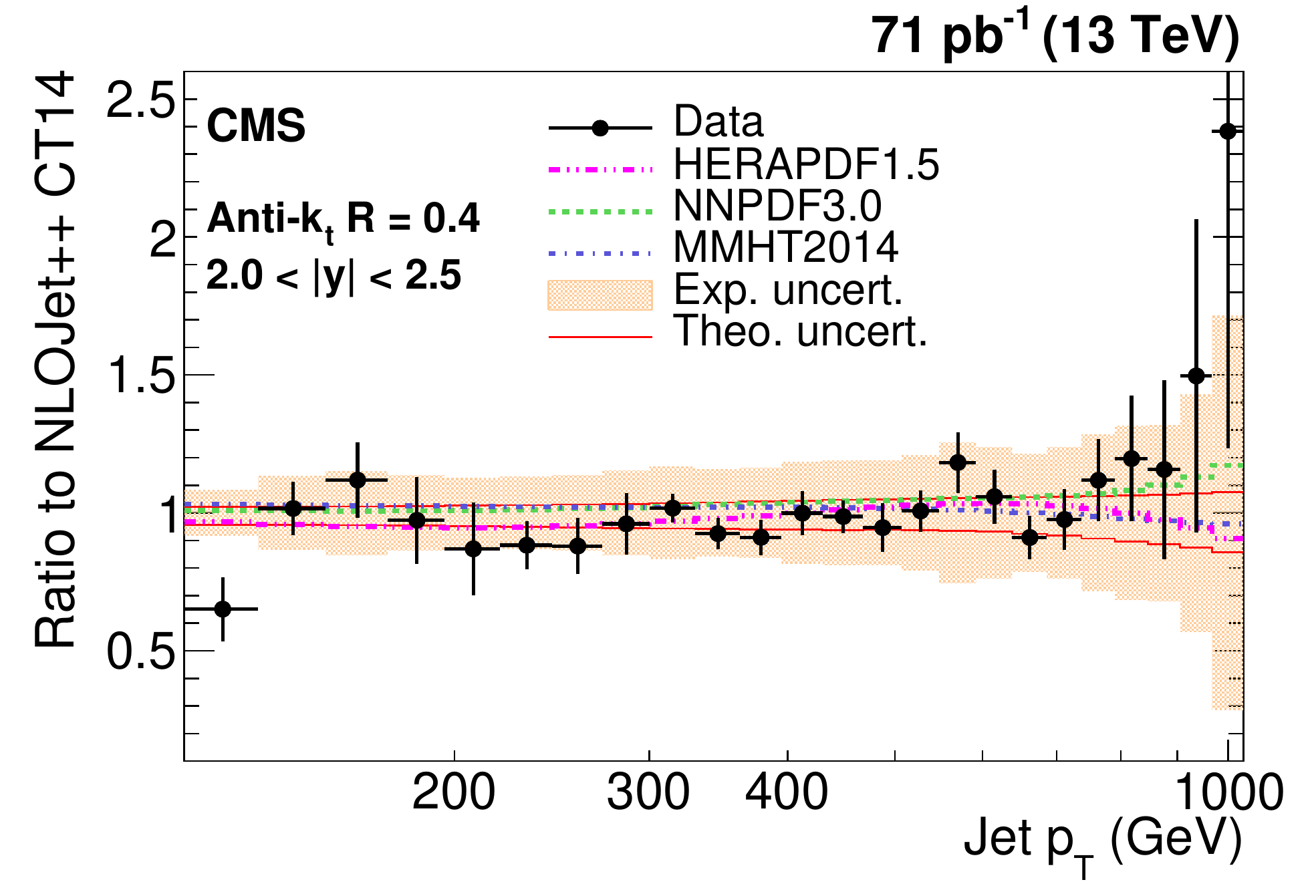}
  \includegraphics[width=0.48\textwidth]{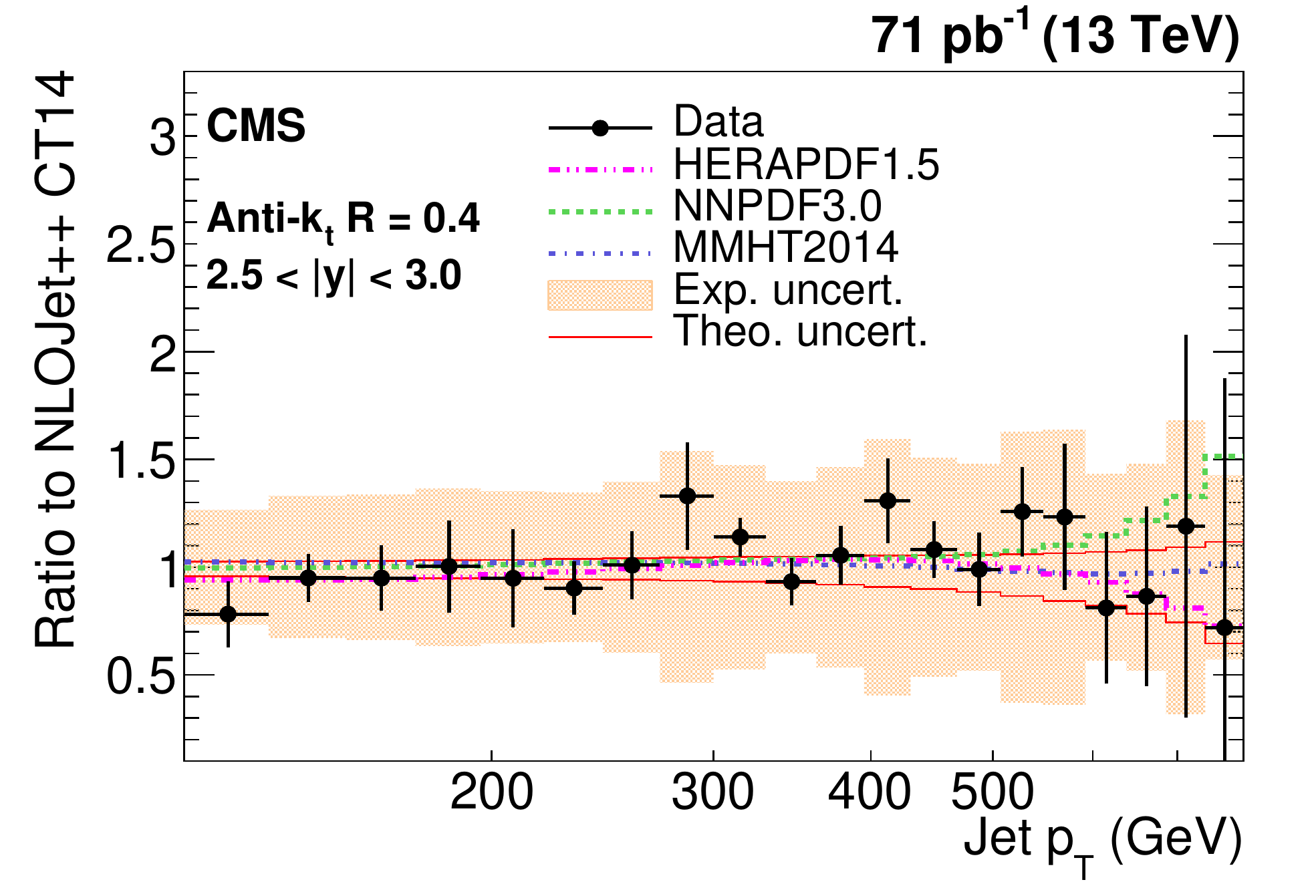}
  \includegraphics[width=0.48\textwidth]{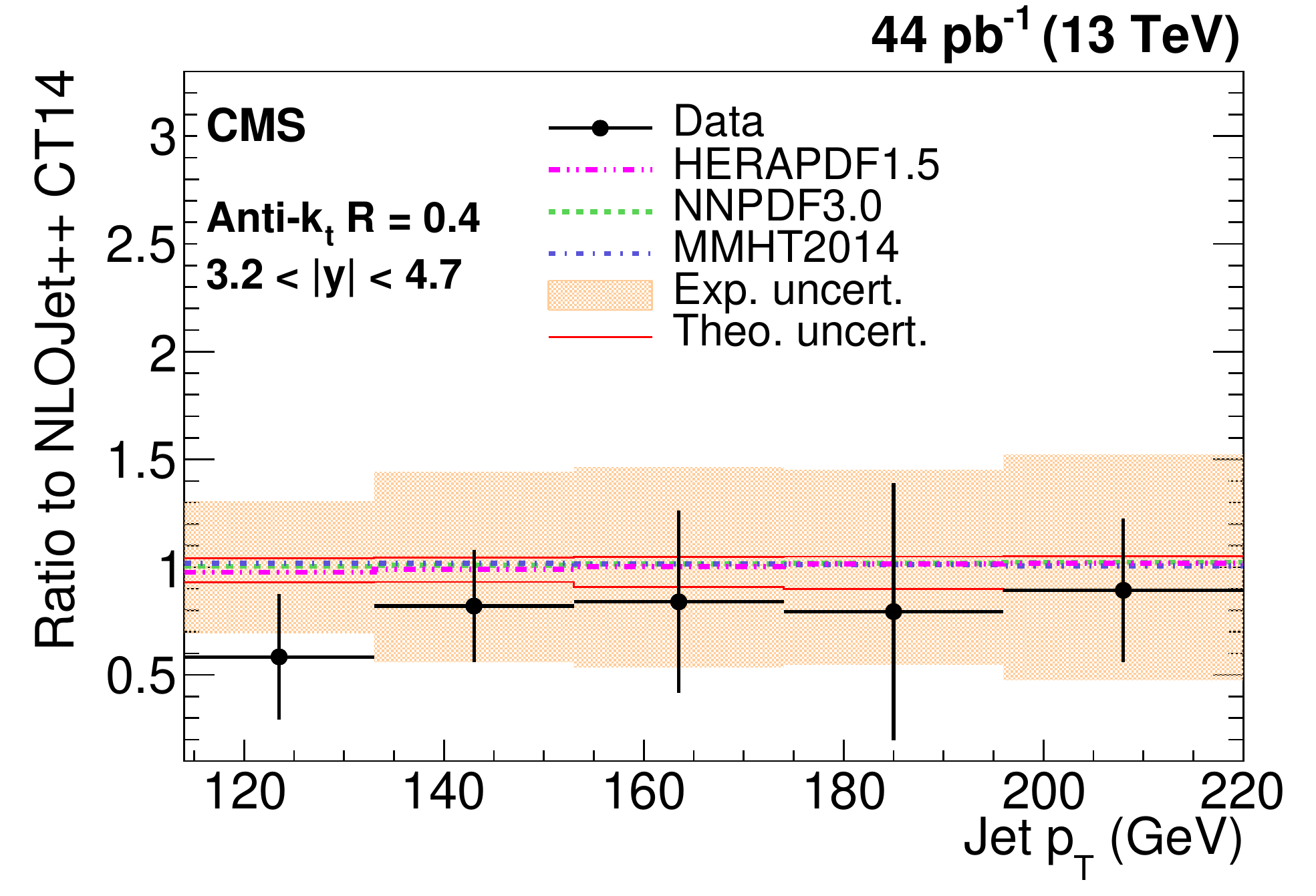}
  \caption{Ratio of measured values to theoretical prediction from
    \NLOJETPP using the CT14 PDF set and corrected for the NP and
    electroweak effects.  Predictions employing three other PDF sets
    are also shown for comparison.  Jets are clustered with the
    anti-\kts algorithm with a distance parameter of 0.4.  The error
    bars correspond to the statistical uncertainties of the data and
    the shaded bands to the total experimental systematic
    uncertainties.}
  \label{fig:AK4_ratio_CT14}
\end{figure*}

The ratios of data over the \NLOJETPP predictions using the CT14 PDF
set are shown in Fig.~\ref{fig:AK7_ratio_CT14} for the AK7 jets.  The
error bars on the points correspond to the statistical uncertainties,
and the shaded bands correspond to the total experimental systematic
uncertainties.  For comparison, predictions employing three
alternative PDF sets are also shown.  Figure~\ref{fig:AK4_ratio_CT14}
shows the results for the AK4 jets.  Overall, a good agreement within
the uncertainties is observed between the data and predictions in the
entire kinematic range studied, for both jet cone sizes.  However, for
$R = 0.4$, the cross sections are systematically overestimated by about 5--10\%,
while a better description is provided for jets reconstructed with $R =
0.7$.  The relatively poor agreement for $R = 0.4$ is due to PS and soft-gluon
resummation contributions, which are missing in fixed-order
calculations, and that are more relevant for smaller jet cone sizes because of out-of-cone effects.

\begin{figure*}[htbp]
  \centering
  \includegraphics[width=0.48\textwidth]{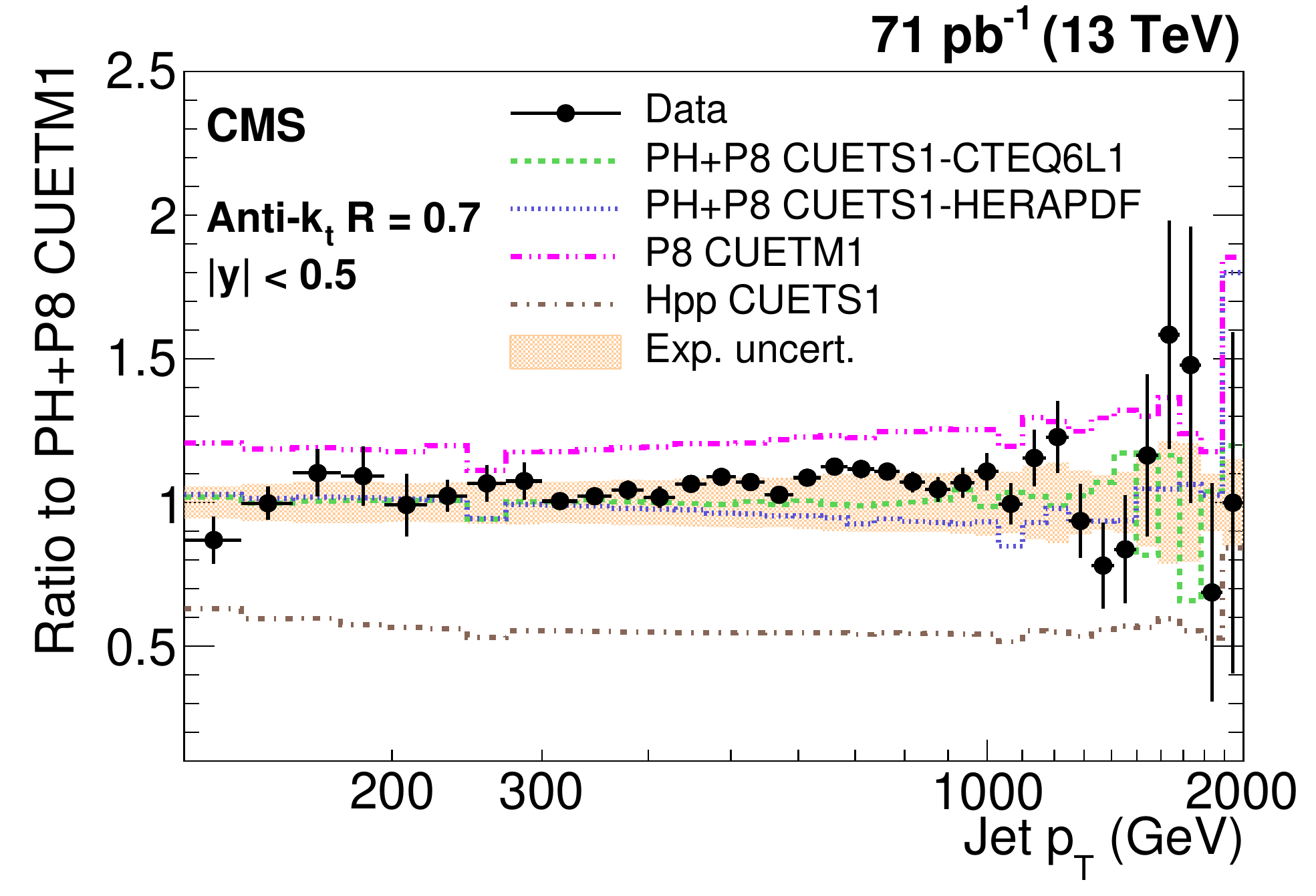}
  \includegraphics[width=0.48\textwidth]{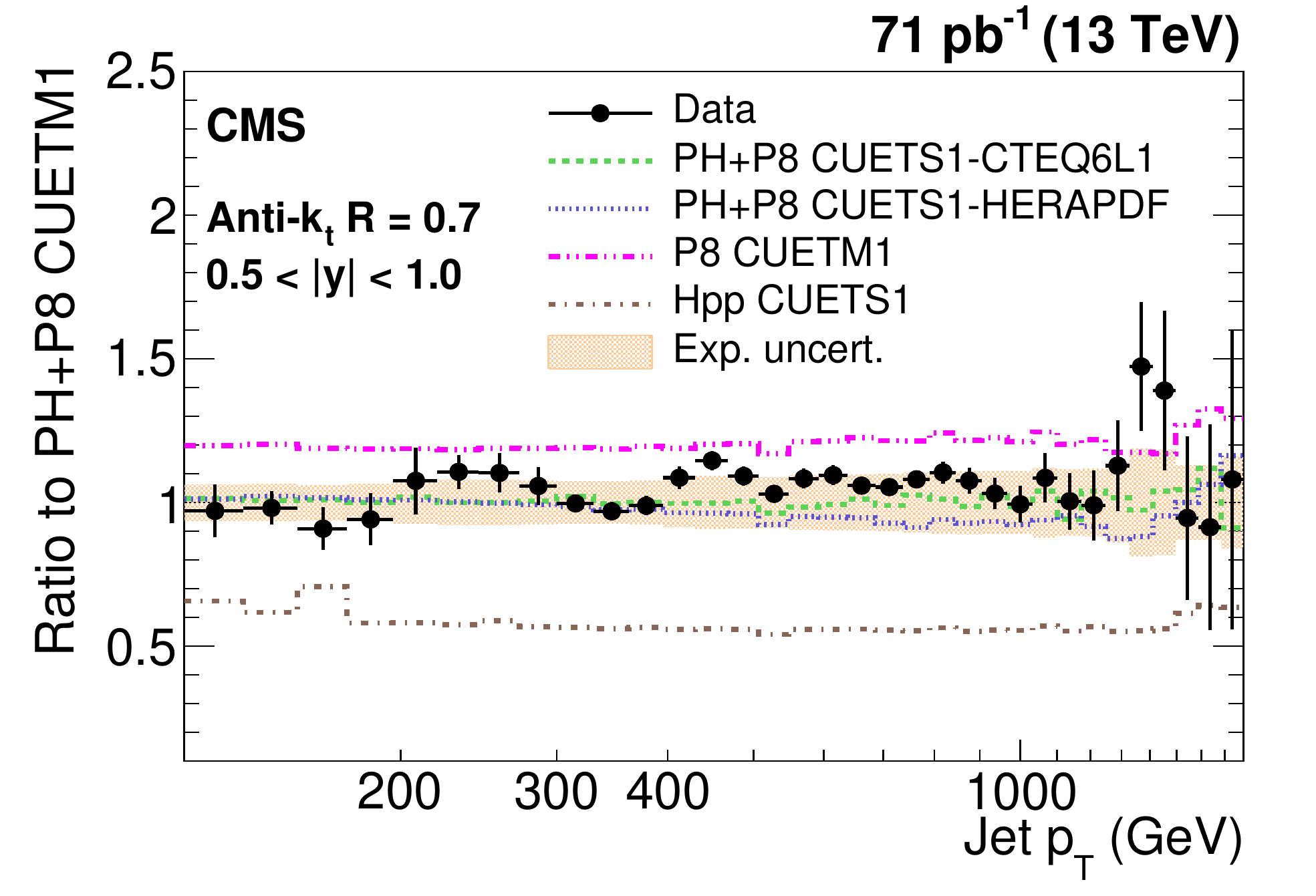}
  \includegraphics[width=0.48\textwidth]{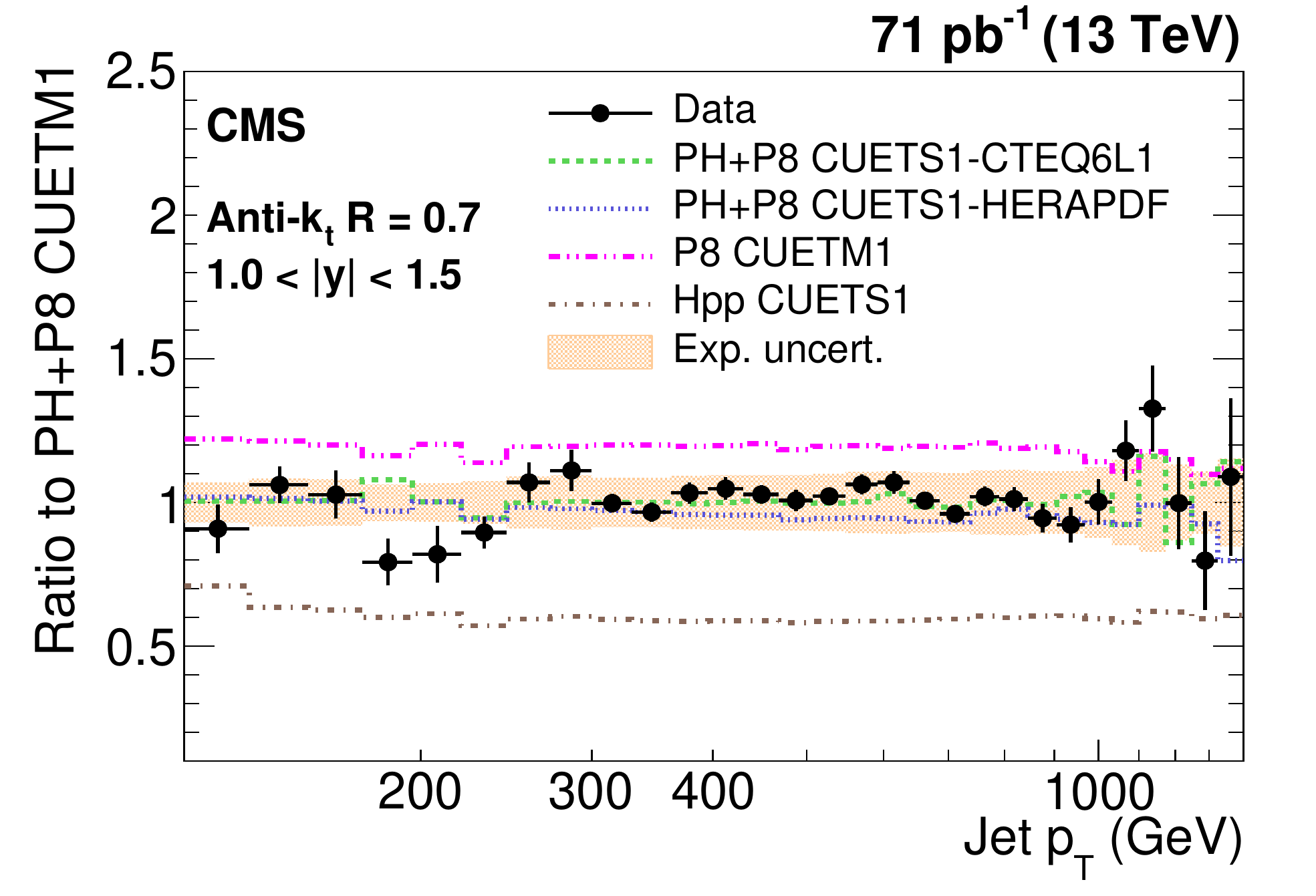}
  \includegraphics[width=0.48\textwidth]{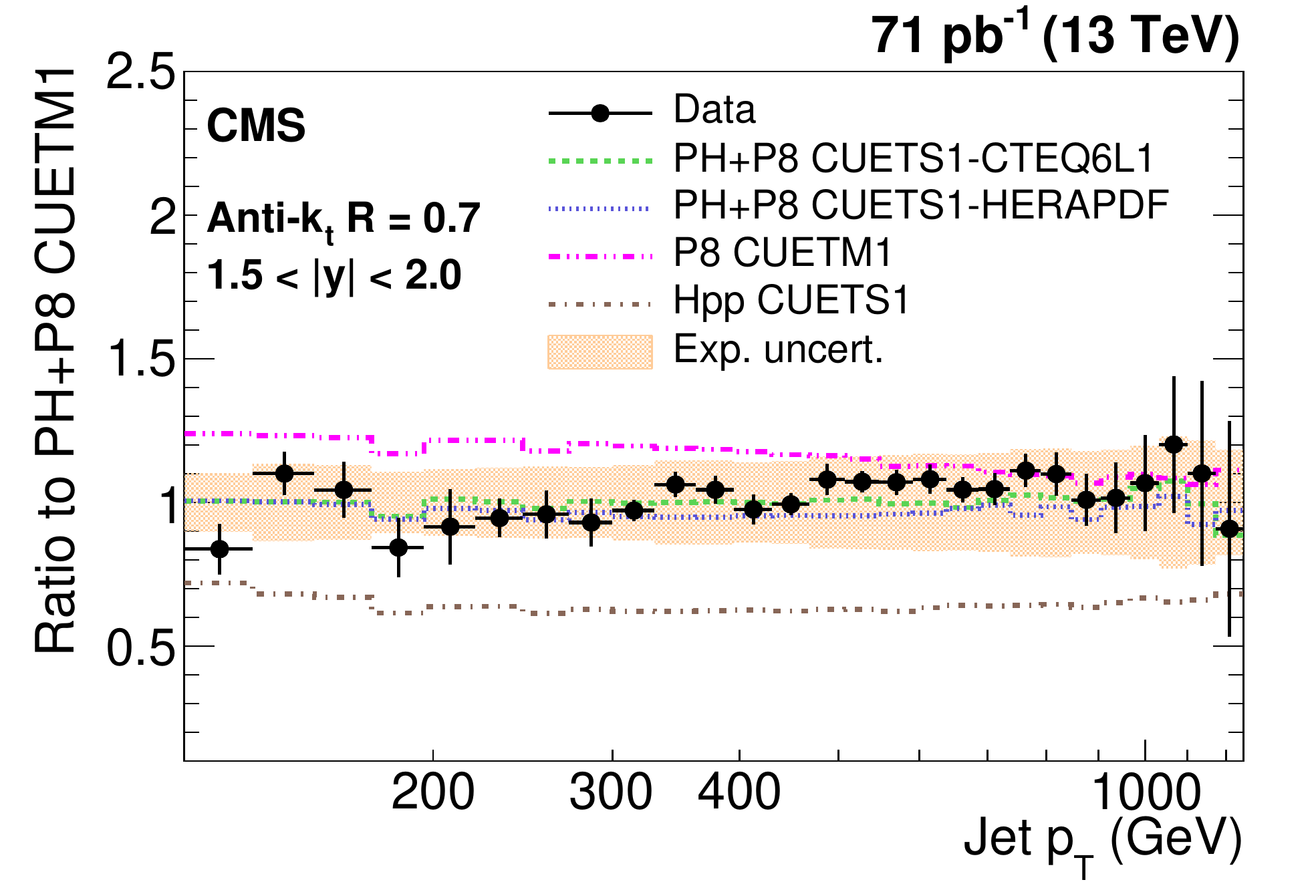}
  \includegraphics[width=0.48\textwidth]{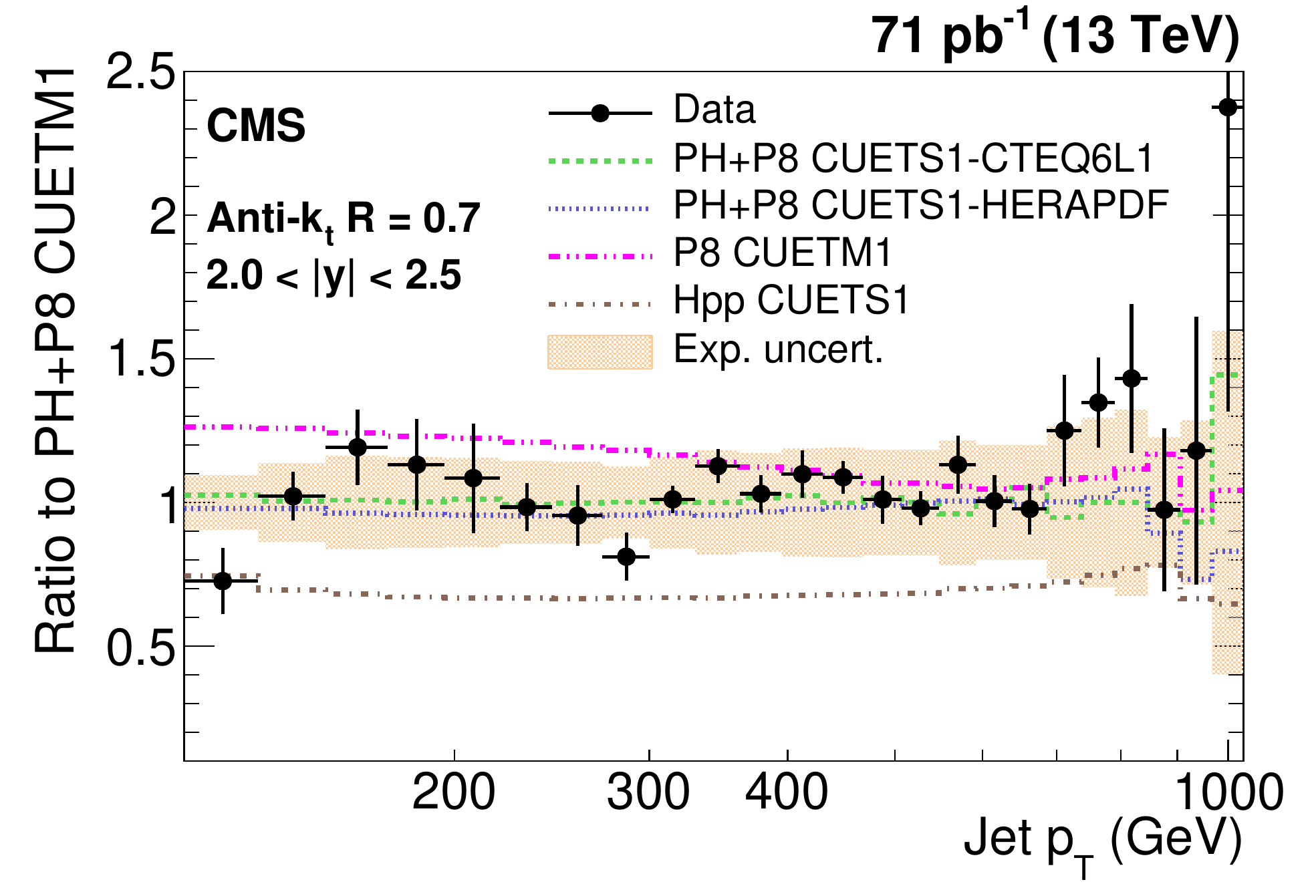}
  \includegraphics[width=0.48\textwidth]{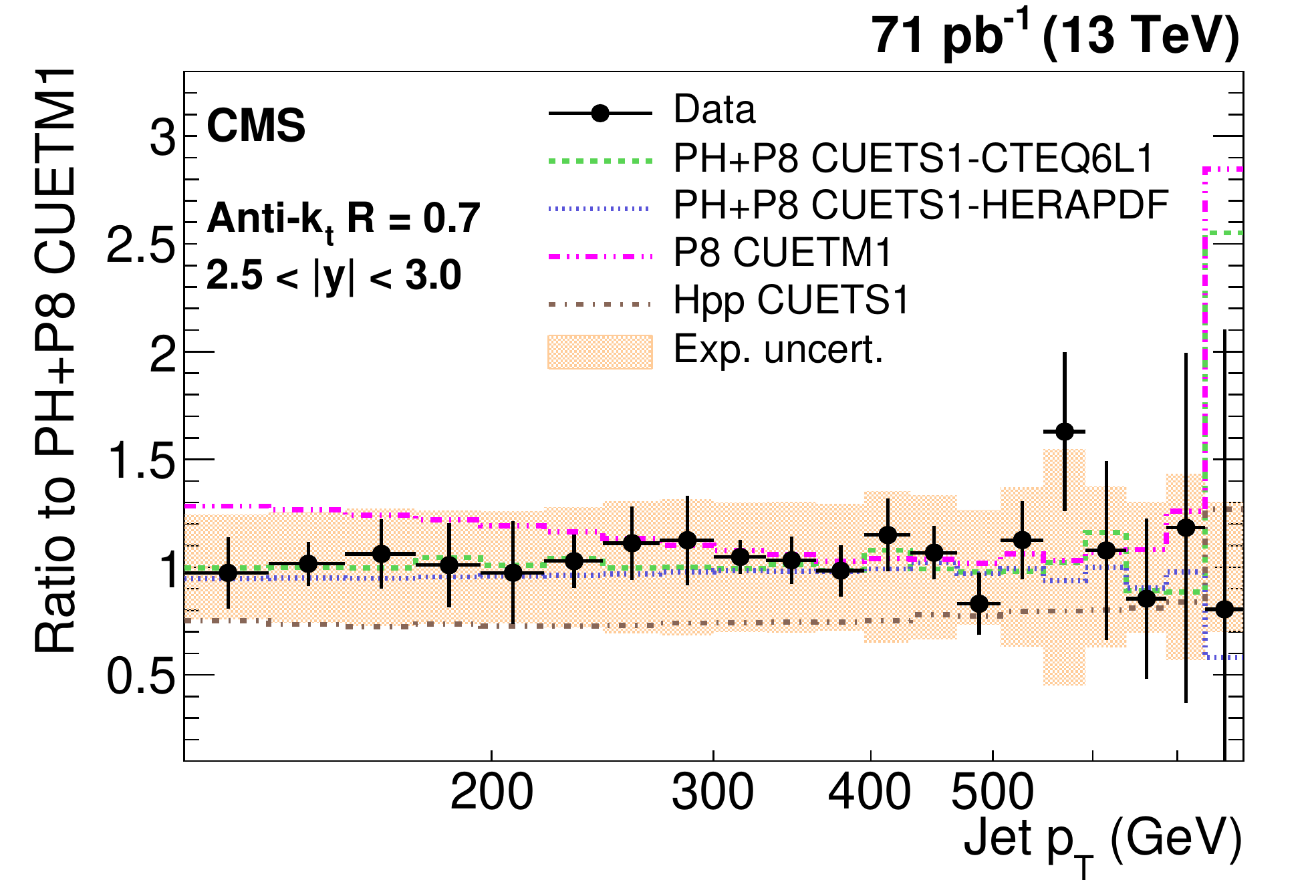}
  \includegraphics[width=0.48\textwidth]{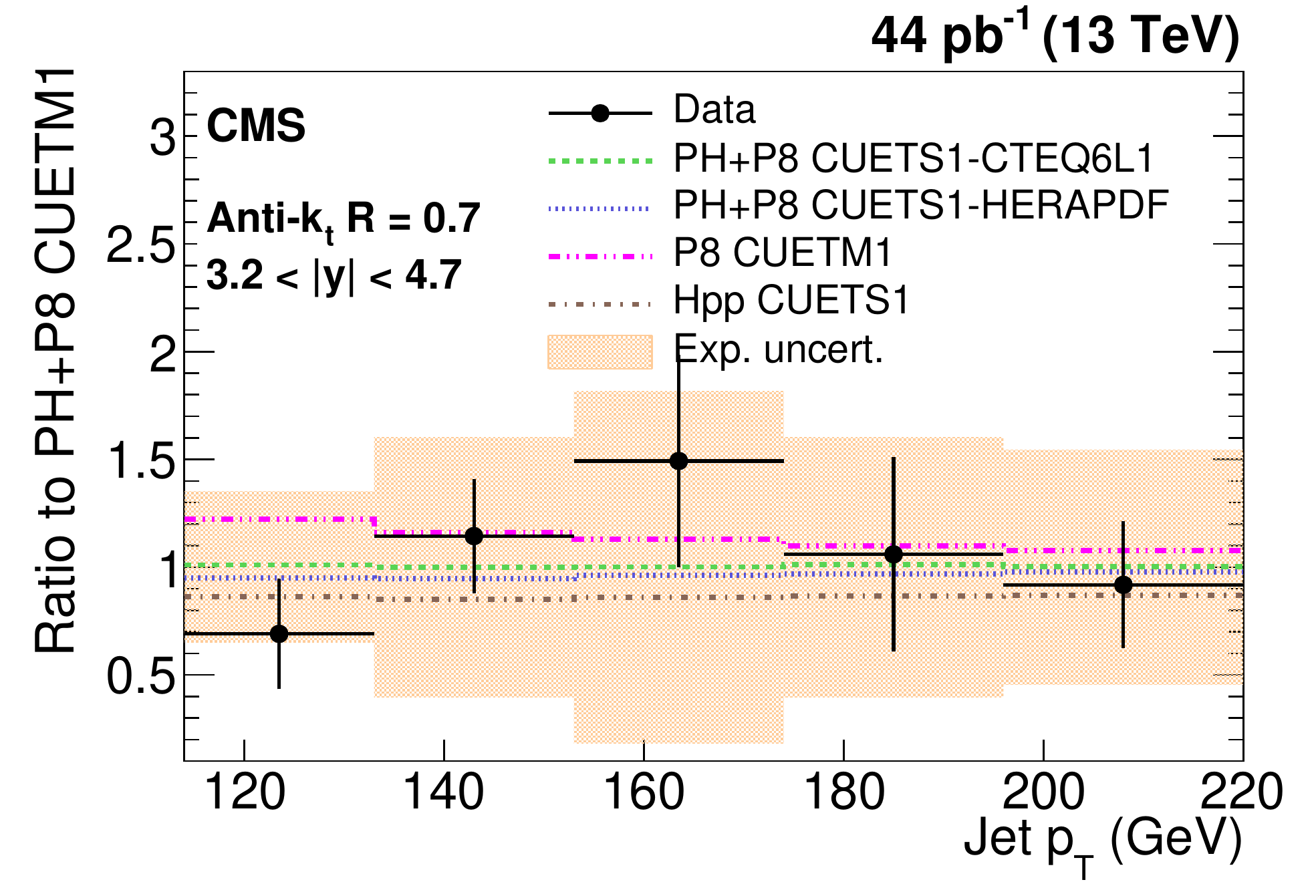}
  \caption{Ratio of measured values to predictions from \POWHEG (PH)
    + \PYTHIAE (P8) with tune CUETM1. Predictions employing four other
    MC generators are also shown for comparison, where PH, P8, and Hpp
    stands for \POWHEG, \PYTHIAE, and \HERWIGpp (HPP),
    respectively. Jets are clustered with the anti-\kts algorithm with
    a distance parameter of 0.7. The error bars correspond to the
    statistical uncertainties of the data and the shaded bands to the
    total experimental systematic uncertainties.}
  \label{fig:AK7_ratio_Powheg}
\end{figure*}

\begin{figure*}[htbp]
  \centering
  \includegraphics[width=0.48\textwidth]{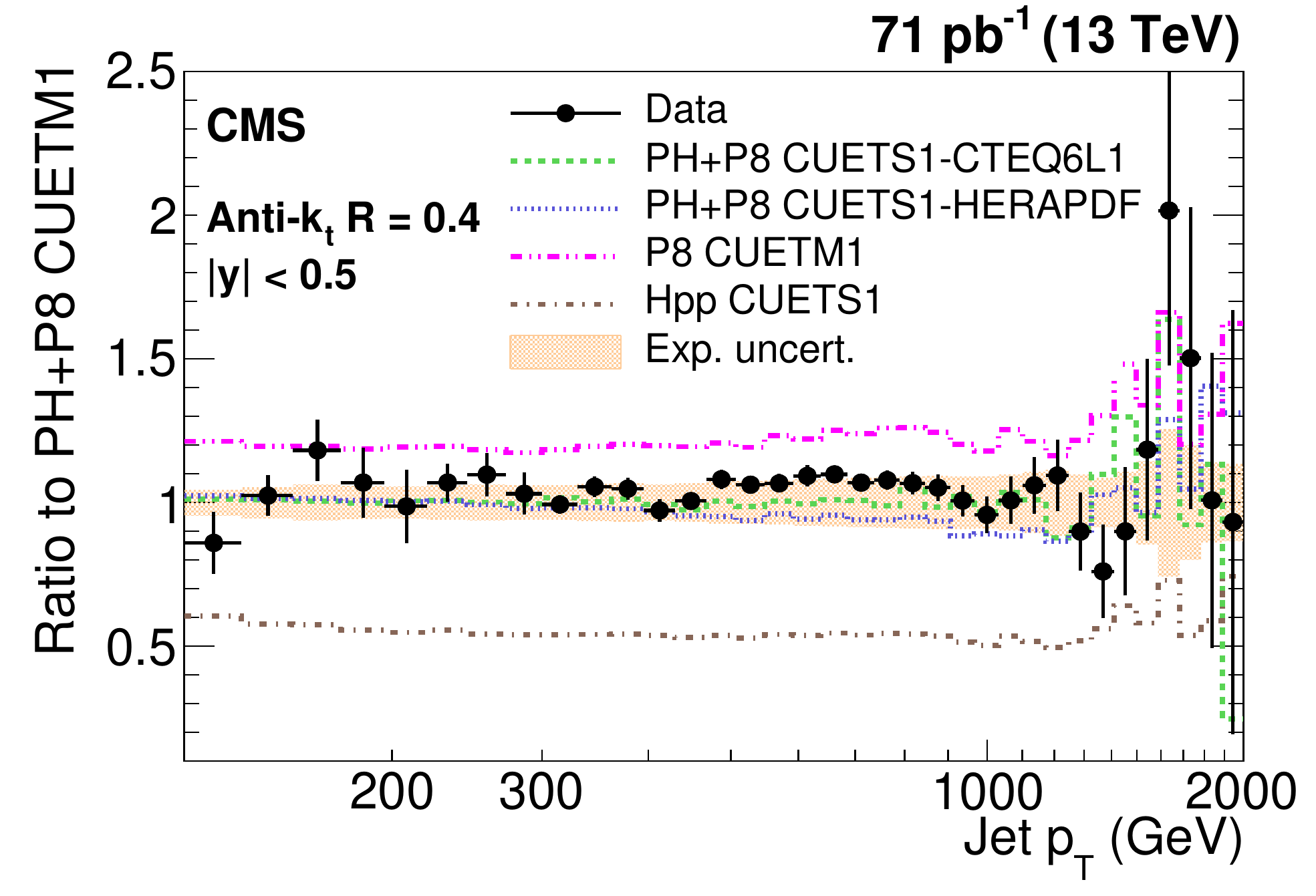}
  \includegraphics[width=0.48\textwidth]{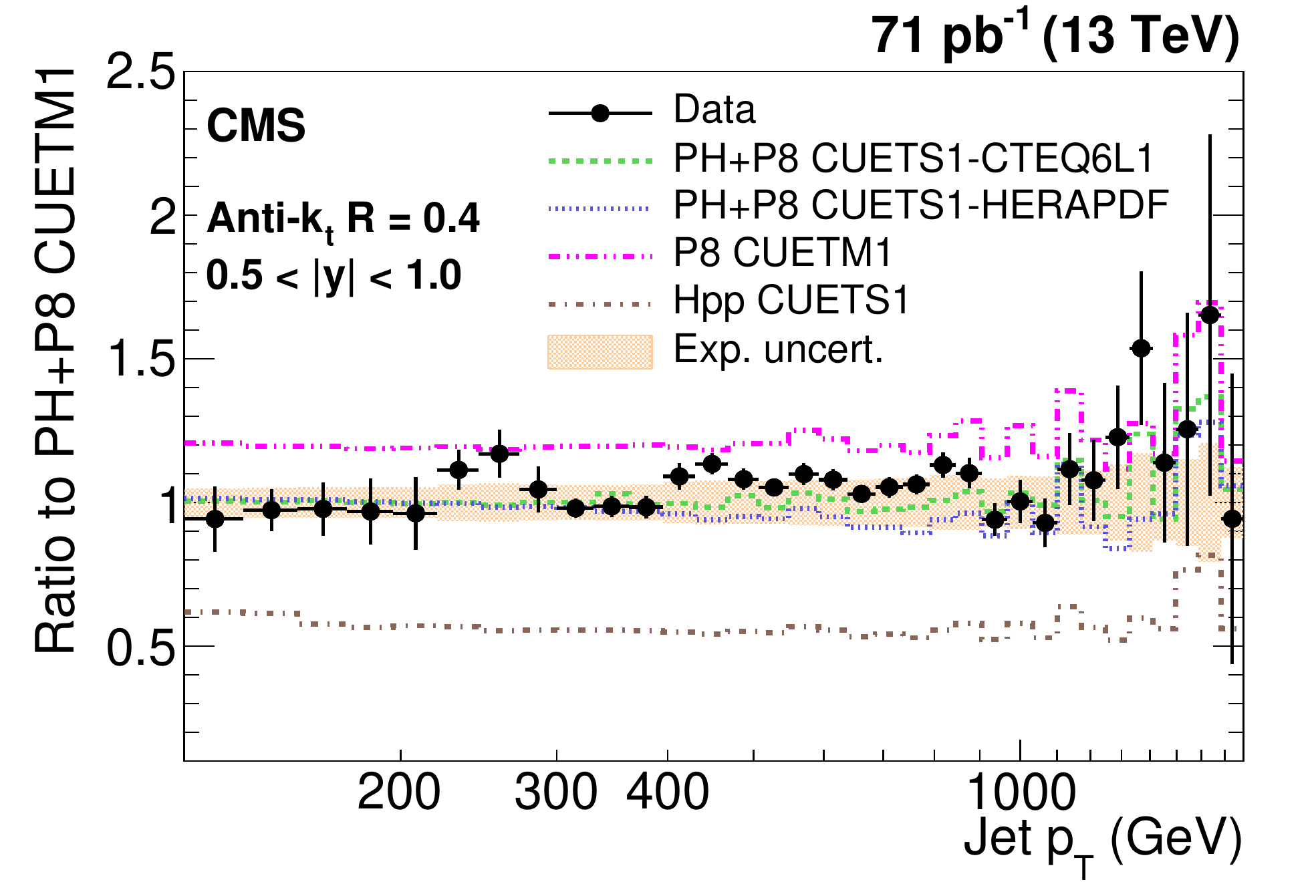}
  \includegraphics[width=0.48\textwidth]{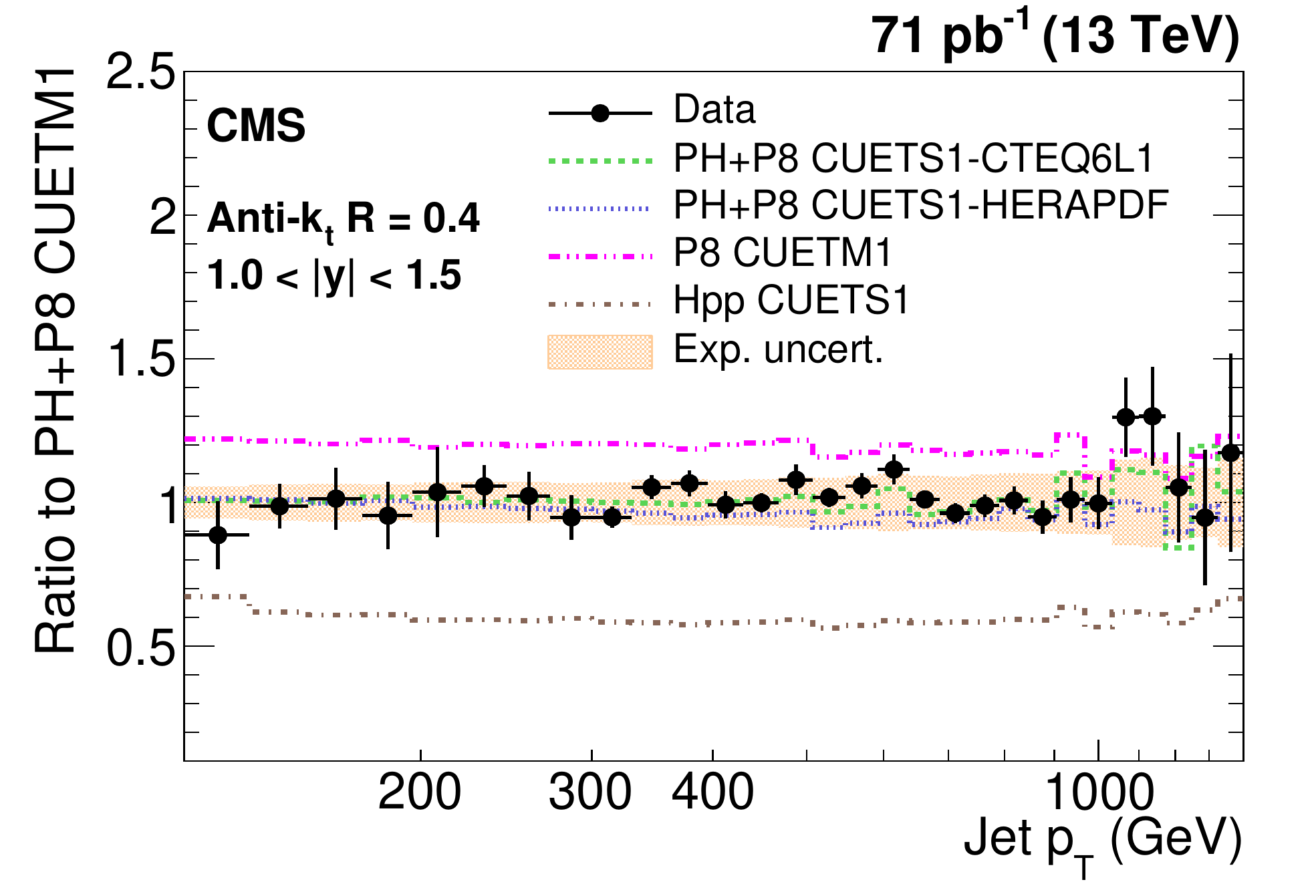}
  \includegraphics[width=0.48\textwidth]{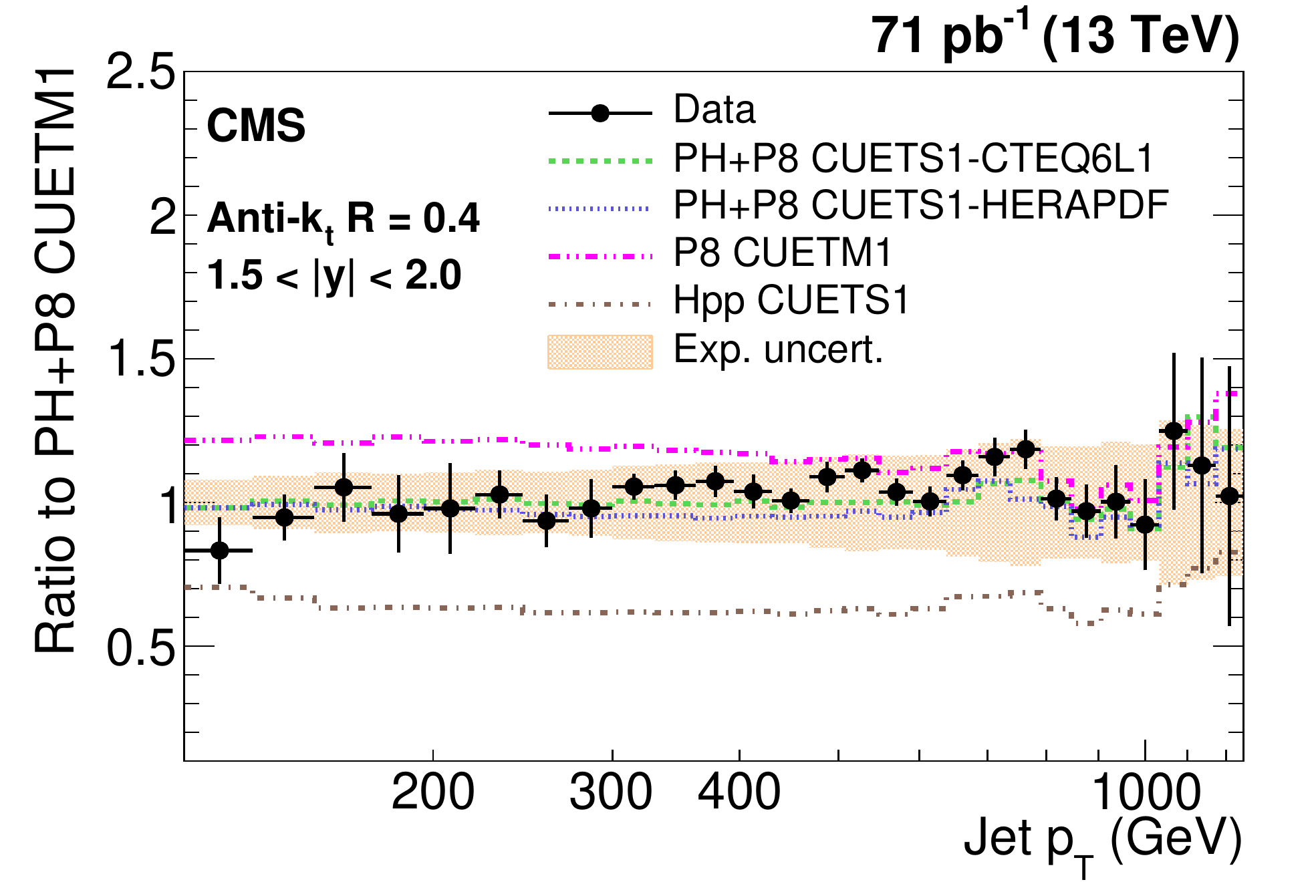}
  \includegraphics[width=0.48\textwidth]{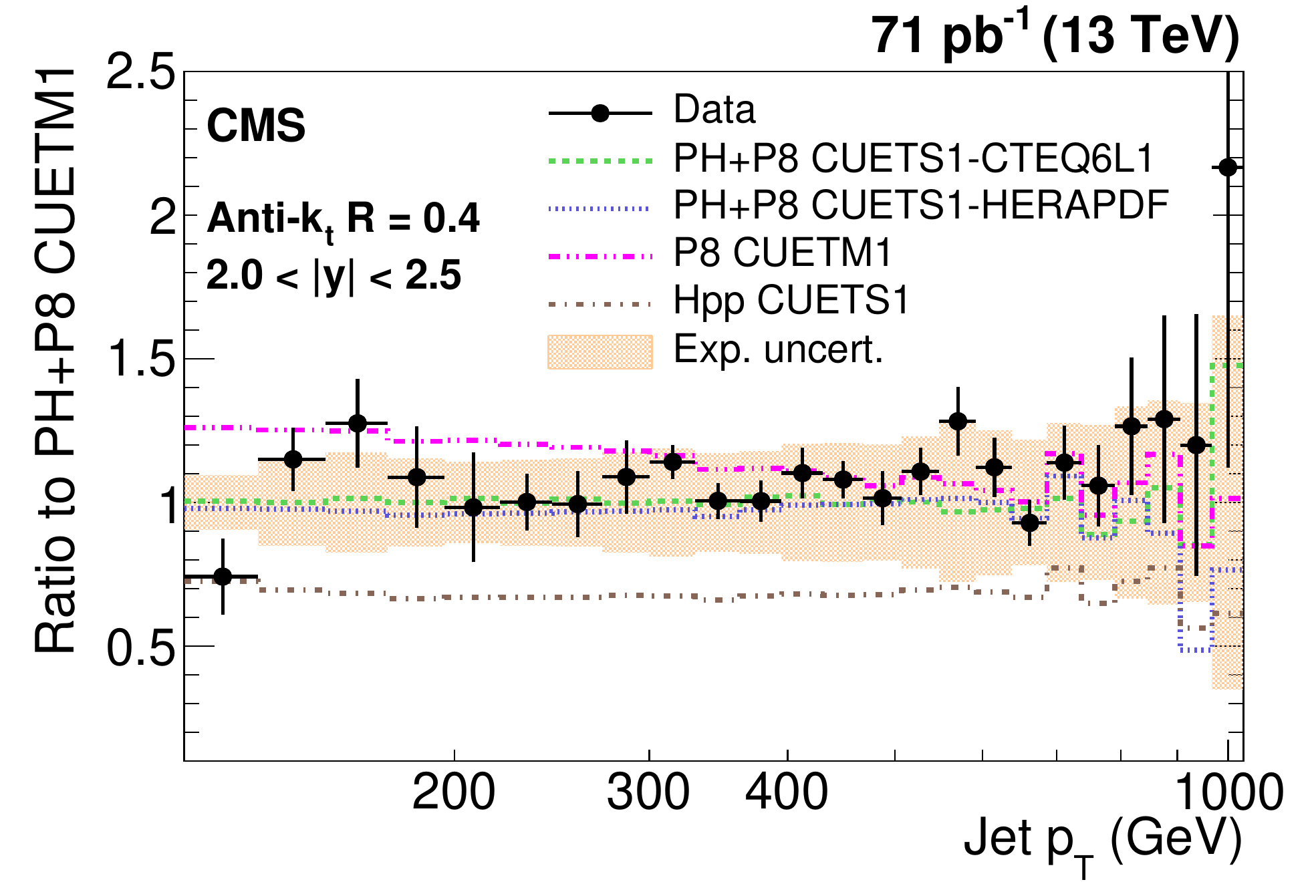}
  \includegraphics[width=0.48\textwidth]{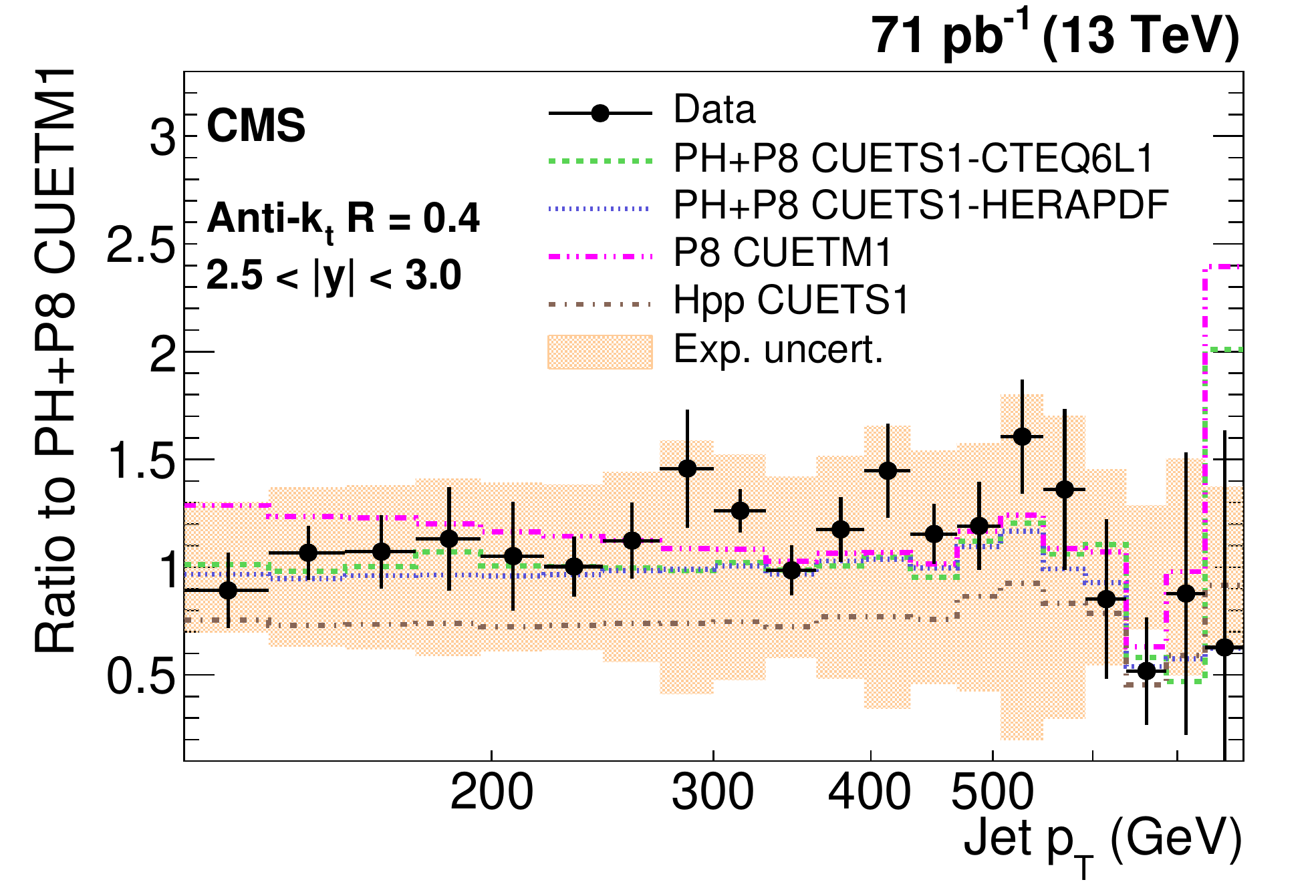}
  \includegraphics[width=0.48\textwidth]{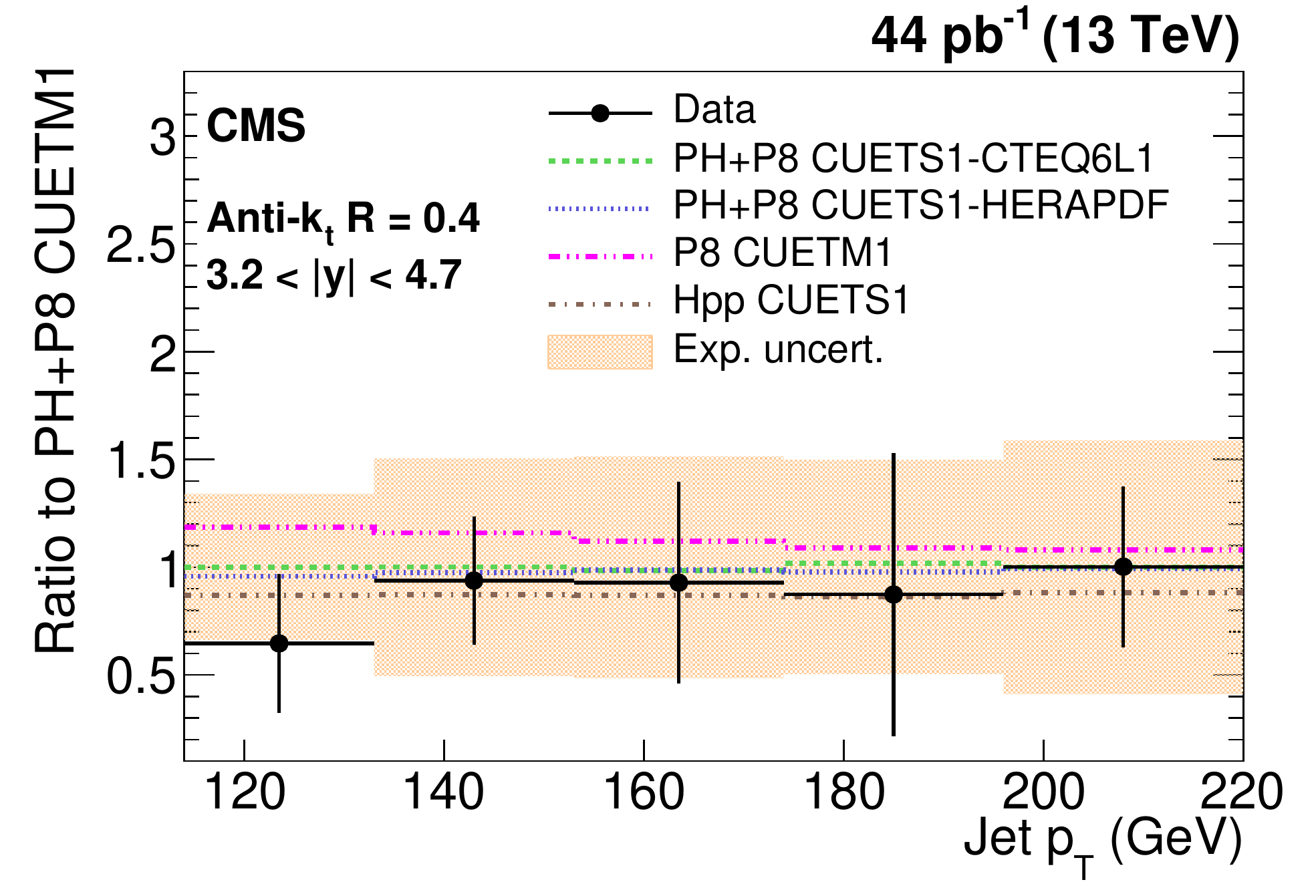}
  \caption{Ratio of measured values to predictions from \POWHEG (PH)
    + \PYTHIAE (P8) with tune CUETM1. Predictions employing four other
    MC generators are also shown for comparison, where PH, P8, and Hpp
    stands for \POWHEG, \PYTHIAE, and \HERWIGpp (HPP),
    respectively. Jets are clustered with the anti-\kts algorithm with
    a distance parameter of 0.4. The error bars correspond to the
    statistical uncertainties of the data and the shaded bands to the
    total experimental systematic uncertainties.}
  \label{fig:AK4_ratio_Powheg}
\end{figure*}

The ratios of data over predictions from \POWHEG + \PYTHIAE with tune
CUETM1 are shown in Figs.~\ref{fig:AK7_ratio_Powheg}(\ref{fig:AK4_ratio_Powheg}) for the AK7(AK4) jets. The error bars
on the points correspond to the statistical uncertainties and the
shaded bands to the total experimental systematic uncertainties. For
comparison, four other MC predictions are also shown. There is an
overall good level of agreement within the uncertainties between data
and predictions from \POWHEG + \PYTHIAE with various tunes for both
cone sizes, in the entire kinematic range studied. The agreement of
data with \PYTHIAE and \HERWIGpp is poor in absolute scale. The \HERWIGpp event generator
shows good agreement with the data in shape for all rapidity bins,
while \PYTHIAE agrees well in shape with the data for only $\abs{y} < 1.5$.

\section{Summary}
\label{sec:summary}

A measurement of the double-differential cross section as a function
of jet \pt and absolute rapidity $\abs{y}$ is presented for two jet
sizes $R=0.4$ and~0.7 using data from proton-proton collisions at
$\sqrt{s}=13\TeV$ collected with the CMS detector. Data samples
corresponding to integrated luminosities of 71\pbinv and 44\pbinv are
used for absolute rapidities $\abs{y}<3$ and for the forward
region $3.2<\abs{y}<4.7$, respectively.

As expected for LO predictions, the MC event generators \PYTHIAE and
\HERWIGpp exhibit significant discrepancies in absolute scale with
respect to data, which are somewhat more pronounced for the case of
\HERWIGpp\@. In contrast, the shape of the inclusive jet \pt
distribution is well described by \HERWIGpp in all rapidity bins.
Predictions from \PYTHIAE start deviating from the observed shape as $\abs{y}$ increases.

In the comparison between data and predictions at NLO in perturbative QCD
including corrections for nonperturbative and electroweak
effects, it is observed that jet cross sections for the larger jet
size of $R=0.7$ are accurately described, while for $R=0.4$ theory
overestimates the cross section by 5--10\% almost globally. In
contrast, NLO predictions matched to parton showers as performed with
\POWHEG + \PYTHIAE for two different tunes, perform equally well for
both jet sizes. This result is consistent with the previous
measurement performed at $\sqrt{s} = 7\TeV$~\cite{Chatrchyan:2014gia},
where it was observed that \POWHEG + \PYTHIAE correctly describes the
$R$ dependence of the inclusive jet cross section, while fixed-order
predictions at NLO were insufficient in that respect.

This measurement is a first indication that jet physics is
as well understood at $\sqrt{s}=13\TeV$ as at smaller
centre-of-mass energies in the phase space accessible with the new
data.

\clearpage\newpage

\begin{acknowledgments}

We would like to thank A.~Huss for providing us with the electroweak
correction factors.
We congratulate our colleagues in the CERN accelerator departments for the excellent performance of the LHC and thank the technical and administrative staffs at CERN and at other CMS institutes for their contributions to the success of the CMS effort. In addition, we gratefully acknowledge the computing centres and personnel of the Worldwide LHC Computing Grid for delivering so effectively the computing infrastructure essential to our analyses. Finally, we acknowledge the enduring support for the construction and operation of the LHC and the CMS detector provided by the following funding agencies: BMWFW and FWF (Austria); FNRS and FWO (Belgium); CNPq, CAPES, FAPERJ, and FAPESP (Brazil); MES (Bulgaria); CERN; CAS, MoST, and NSFC (China); COLCIENCIAS (Colombia); MSES and CSF (Croatia); RPF (Cyprus); MoER, ERC IUT and ERDF (Estonia); Academy of Finland, MEC, and HIP (Finland); CEA and CNRS/IN2P3 (France); BMBF, DFG, and HGF (Germany); GSRT (Greece); OTKA and NIH (Hungary); DAE and DST (India); IPM (Iran); SFI (Ireland); INFN (Italy); MSIP and NRF (Republic of Korea); LAS (Lithuania); MOE and UM (Malaysia); BUAP, CINVESTAV, CONACYT, LNS, SEP, and UASLP-FAI (Mexico); MBIE (New Zealand); PAEC (Pakistan); MSHE and NSC (Poland); FCT (Portugal); JINR (Dubna); MON, RosAtom, RAS and RFBR (Russia); MESTD (Serbia); SEIDI and CPAN (Spain); Swiss Funding Agencies (Switzerland); MST (Taipei); ThEPCenter, IPST, STAR and NSTDA (Thailand); TUBITAK and TAEK (Turkey); NASU and SFFR (Ukraine); STFC (United Kingdom); DOE and NSF (USA).

Individuals have received support from the Marie-Curie programme and the European Research Council and EPLANET (European Union); the Leventis Foundation; the A. P. Sloan Foundation; the Alexander von Humboldt Foundation; the Belgian Federal Science Policy Office; the Fonds pour la Formation \`a la Recherche dans l'Industrie et dans l'Agriculture (FRIA-Belgium); the Agentschap voor Innovatie door Wetenschap en Technologie (IWT-Belgium); the Ministry of Education, Youth and Sports (MEYS) of the Czech Republic; the Council of Science and Industrial Research, India; the HOMING PLUS programme of the Foundation for Polish Science, cofinanced from European Union, Regional Development Fund; the Mobility Plus programme of the Ministry of Science and Higher Education (Poland); the OPUS programme of the National Science Center (Poland); the Thalis and Aristeia programmes cofinanced by EU-ESF and the Greek NSRF; the National Priorities Research Program by Qatar National Research Fund; the Programa Clar\'in-COFUND del Principado de Asturias; the Rachadapisek Sompot Fund for Postdoctoral Fellowship, Chulalongkorn University (Thailand); the Chulalongkorn Academic into Its 2nd Century Project Advancement Project (Thailand); and the Welch Foundation, contract C-1845.

\end{acknowledgments}
\bibliography{auto_generated}
\cleardoublepage \appendix\section{The CMS Collaboration \label{app:collab}}\begin{sloppypar}\hyphenpenalty=5000\widowpenalty=500\clubpenalty=5000\textbf{Yerevan Physics Institute,  Yerevan,  Armenia}\\*[0pt]
V.~Khachatryan, A.M.~Sirunyan, A.~Tumasyan
\vskip\cmsinstskip
\textbf{Institut f\"{u}r Hochenergiephysik der OeAW,  Wien,  Austria}\\*[0pt]
W.~Adam, E.~Asilar, T.~Bergauer, J.~Brandstetter, E.~Brondolin, M.~Dragicevic, J.~Er\"{o}, M.~Flechl, M.~Friedl, R.~Fr\"{u}hwirth\cmsAuthorMark{1}, V.M.~Ghete, C.~Hartl, N.~H\"{o}rmann, J.~Hrubec, M.~Jeitler\cmsAuthorMark{1}, A.~K\"{o}nig, I.~Kr\"{a}tschmer, D.~Liko, T.~Matsushita, I.~Mikulec, D.~Rabady, N.~Rad, B.~Rahbaran, H.~Rohringer, J.~Schieck\cmsAuthorMark{1}, J.~Strauss, W.~Treberer-Treberspurg, W.~Waltenberger, C.-E.~Wulz\cmsAuthorMark{1}
% ...
\vskip\cmsinstskip
% ...
1:~~Also at Vienna University of Technology, Vienna, Austria\\
% ...

\end{sloppypar}
\end{document}